\documentclass[prd,aps,showpacs,superscriptaddress,nofootinbib,amsmath,amssymb]{revtex4-1}
\usepackage{graphicx}
\usepackage{dcolumn}
\usepackage{bm}
\usepackage{lipsum}
\usepackage{multirow}
\usepackage{adjustbox}
\graphicspath{{./eps/}}

\begin{document}

\title{Quasielastic production of polarized hyperons in antineutrino--nucleon reactions}

\def\a              {\alpha}
\def\b              {\beta}
\def\m              {\mu}
\def\n              {\nu}
\def\ti             {\tilde}
\def\e              {\varepsilon}
\def\r              {\rho}
\def\s              {\sigma}
\def\g              {\gamma}
\def\d              {\delta}
\def\t              {\tau}

\author{F. \surname{Akbar}}

\author{M. Rafi \surname{Alam}}

\author{M. Sajjad \surname{Athar}}
\email{sajathar@gmail.com}

\author{S. K. \surname{Singh}}
\affiliation{Department of Physics, Aligarh Muslim University, Aligarh-202002, India}

\begin{abstract} 
We have studied the differential cross section as well as the longitudinal and perpendicular components of polarization of final 
hyperon($\Lambda$,$\Sigma$) produced in the antineutrino induced quasielastic charged 
current reactions on nucleon and nuclear targets.
The nucleon-hyperon transition form factors are determined from the
experimental data on quasielastic $(\Delta S =0)$ charged current (anti)neutrino--nucleon 
scattering and the semileptonic decay of neutron and hyperons  assuming 
G--invariance, T--invariance and SU(3) symmetry.
The vector transition form factors are obtained in terms of nucleon 
electromagnetic form factors for which various parameterizations available in 
literature have been used. 
 A dipole parameterization for the axial vector form factor and the pseudoscalar transition form factor 
derived in terms of axial vector form factor assuming PCAC and GT relation 
extended to strangeness sector have been used in numerical evaluations.
 The flux averaged cross section and polarization observables corresponding to CERN Gargamelle
experiment have been calculated for quasielastic hyperon production and found to be in reasonable
agreement with the experimental observations. The numerical results
for the flux averaged differential cross section$\frac{d\sigma}{dQ^2}$ and 
longitudinal(perpendicular) polarization $P_L(Q^2)(P_P(Q^2))$ 
relevant for the  antineutrino fluxes of MINER$\nu$A,
MicroBooNE, and T2K experiments have been presented. This will be 
useful in interpreting future experimental results on production cross sections and 
polarization observables from the experiments on the quasielastic production of hyperons induced by antineutrinos
 and explore the possibility of determining the axial vector and  pseudoscalar form factors in the strangeness sector.
\end{abstract}
\maketitle
\section{Introduction}
Our knowledge of the transition form factors in the antineutrino induced quasielastic process of hyperon production ($|\Delta S| = 1$)  is far from satisfactory. 
Recently, with the development of high intensity (anti)neutrino beams in the few GeV region, considerable 
 interest has developed in understanding these weak transition 
form factors specially in the axial vector sector.
  These form factors have
been determined experimentally and theoretically using Cabibbo theory assuming SU(3) symmetry and other
symmetries of weak hadronic currents in the Standard Model. Most of these  form factors are determined from the analysis of 
semileptonic decay of hyperons and neutron which are limited to very low momentum transfer.  These form factors are found to be consistent with SU(3) symmetry 
which relates them to the form factors in $\Delta S =0$ sector of (anti)neutrino--nucleon scattering
and to the various couplings in semileptonic hyperon decays. However, the status of G--invariance, conservation of vector current(CVC),
partial conservation of axial current(PCAC), etc.
which seem to work quite well in the nucleon sector are not well understood when extended
to octet of baryons using SU(3) symmetry which is known to be an approximate symmetry. Even though, the vast amount of data available on semileptonic
 decay of hyperons is consistent with the assumption of SU(3) symmetry, the violation of G--invariance and SU(3) symmetry is not ruled out~\cite{Cabibbo:2003cu}.
There is no unambiguous way to implement SU(3) symmetry as far as CVC and PCAC are concerned but the prescriptions which have been used in
literature to implement the symmetry, seem to work well~\cite{Cabibbo:2003cu,Gaillard:1984ny,Gazia}.

  The charged current quasielastic production of hyperons by antineutrinos (charged current quasielastic production induced by neutrinos is prohibited by
$\Delta S = \Delta Q$ rule while any neutral current production induced by $\nu$ and $\bar\nu$ is prohibited by the absence of Flavor Changing Neutral Current(FCNC) in the Standard Model) is the most 
appropriate place to study the nucleon-hyperon transition
form factors which enables us to extend the study of form factors  to higher $Q^2$ 
beyond the $Q^2$ values accessible in semileptonic hyperon decays.
There are some experimental studies performed to determine these form factors from the cross section measurements done for these processes at
CERN~\cite{Erriquez:1977tr,Eichten:1972bb,Erriquez:1978pg}, BNL~\cite{Fanourakis:1980si}, FNAL~\cite{Ammosov:1986jn,Ammosov:1986xv} and 
Serpukhov~\cite{Brunner:1989kw} which are limited by low statistics.  Theoretically, these reactions have been studied for more than 50
years~\cite{Bell:1962,Chilton:1964zza,Adler:1965,Egart:1963nc,Cabibbo:1965zza,Pais:1971er,Marshak,LlewellynSmith:1971uhs,Finjord:1975zy,Block:1964gj,sirlin:1965} 
but recently there has been renewed interest in studying these reactions~\cite{Singh:2006xp,Mintz:2007zz,Mintz:2006yp,Kuzmin:2008zz,Wu:2013kla,Alam:2014bya}
due to the feasibility of doing experiments with the availability 
of high intensity antineutrino
beams~\cite{fermi,JPARC,Chen:2007ae,Palamara:2011zz,Acciarri:2015uup,Fields:2013zhk}. Most of the theoretical calculations have been done only for the production cross section but there 
exist some calculations also for the polarization of the produced hyperons~\cite{Adler:1965,Egart:1963nc,Cabibbo:1965zza,Pais:1971er,Marshak,LlewellynSmith:1971uhs}. 
There is only one experiment done at CERN which has reported the results for the polarization observables
for $\Lambda$ hyperon produced in the quasielastic $\bar \nu_\mu p \to \mu^+ \Lambda$ reaction~\cite{Erriquez:1978pg}.

Experimentally, there is now possibility to study the production cross section of hyperons and other strange particles as well as polarization of hyperons at present
facilities at Fermilab~\cite{fermi} 
and J-PARC~\cite{JPARC} where high intensity beams of (anti)neutrino are available. The experiments planned with liquid argon TPC (LArTPC) detectors at
MicroBooNE~\cite{Chen:2007ae}, 
and ArgoNeuT~\cite{Palamara:2011zz}, and the proposed DUNE~\cite{Acciarri:2015uup} and  LAr1-ND, ICARUS-T600~\cite{combined}
experiments at Fermilab will be able to see charged hadrons in coincidence, thus making it possible to measure polarization in addition to the 
cross section measurements being done at MINER$\nu$A~\cite{Fields:2013zhk}. It is, therefore, most 
appropriate time to theoretically perform the calculations for the polarization observables in addition to the differential cross sections in the
 Standard Model using Cabibbo  theory and/or quark models, 
using the present state of knowledge
 about the symmetry of weak hadronic currents and the properties of transition form factors associated with the matrix element between the hadronic states.
 Since these experiments are
 planned to be performed using nuclear targets, it is important that we understand the implications of nuclear medium effects in the interpretation of the
 experimental results.
 This will facilitate the analysis of experimental results when they become available. 
 We propose to study theoretically the production and polarization of hyperons produced
in the following reactions:
\begin{eqnarray}\label{reaction}
\bar \nu_\mu + p &\longrightarrow& \mu^+ + \Lambda \nonumber \\
\bar \nu_\mu + p &\longrightarrow& \mu^+ + \Sigma^0 \nonumber \\
\bar \nu_\mu + n &\longrightarrow& \mu^+ + \Sigma^-, 
\end{eqnarray}
 on nucleons and nuclear targets using Cabibbo theory in the Standard Model with GIM mechanism for extension to strangeness sector. We also assume the T--invariance and the absence of second
class currents which forbid the existence of hyperon polarization perpendicular to the reaction plane.

In section-\ref{Formalism}, we describe in brief the formalism for calculating the cross section and polarization of hyperons produced in
 the quasielastic antineutrino reactions on free and bound nucleons. 
The effect of nuclear medium arising due to Fermi motion and Pauli blocking of initial nucleon states are also considered. We have in this paper 
 not taken into account the final state interaction effect of outgoing polarized hyperons, the work for which 
  is in progress and will be reported elsewhere. In section-\ref{Results and Discussion}, we present the results and discussion, and give summary and conclusions in section-\ref{sec:summary}.
 \section{Formalism}\label{Formalism}
 \begin{figure}
 \begin{center}
    \includegraphics[height=3cm,width=8cm]{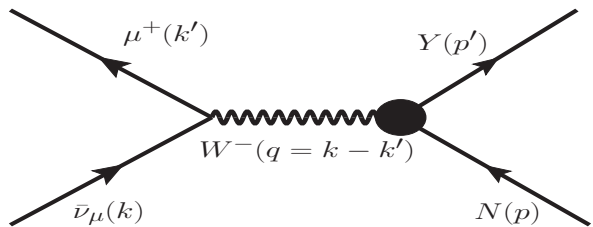}
  \caption{Feynman diagram  for the process $\bar{\nu}_\mu(k) + N(p)\rightarrow \mu^+(k^\prime) + Y(p^\prime)$, 
  where $N$ and $Y$ stand for initial nucleon and final hyperon,
  respectively. 
  The quantities in the bracket represent 
  four momentum of the corresponding particles.}\label{fyn_hyp}
   \end{center}
 \end{figure}
\subsection{Matrix element and transition form factors}
   The transition matrix element for the process 
 \begin{equation}\label{hyp-rec}
 \bar{\nu}_\mu(k) + N(p) \rightarrow \mu^+(k^\prime) + Y(p^\prime),~~(N~=~p,~n;~Y~=~\Lambda,~\Sigma)  \nonumber \\ 
 \end{equation}
depicted in Fig.~\ref{fyn_hyp}, 
is written as  
 \begin{equation}\label{matrix1}
 {\cal{M}}=\frac{G_F}{\sqrt{2}}\sin \theta_c\, l^\mu~\left[\bar u_Y(p^\prime) J_\mu u_N(p)\right].
 \end{equation}
  In the above expression $G_F$ is the Fermi coupling constant and $\theta_c$ is the Cabibbo angle. 
  $l^\mu$ is the leptonic current given by
  \begin{equation}\label{lep}
  l^\mu= \bar u(k^\prime) \gamma^\mu(1+\gamma_5) u(k),
  \end{equation}
 and  $J_{\mu}$ is the hadronic current operator given by
  \begin{equation}\label{had}
  J_{\mu} = V_{\mu} - A_{\mu} 
 \end{equation}
 where
 \begin{eqnarray}\label{vec}
  V_{\mu} &=& \gamma_\mu f^{NY}_1(Q^2)+i\sigma_{\mu\nu} 
 \frac{q^\nu}{m_N+m_Y} f^{NY}_2(Q^2) + \frac{q_\mu}{m_N+m_Y} f^{NY}_3(Q^2) 
 \end{eqnarray}
 and
 \begin{eqnarray}\label{axi}
  A_{\mu} &=& \gamma_\mu \gamma_5 g^{NY}_1(Q^2) + 
  i \sigma_{\mu\nu}\gamma_5 \frac{q^\nu}{m_N+m_Y} g^{NY}_2(Q^2)+ \frac{q_\mu} {m_N+m_Y} g^{NY}_3(Q^2) \gamma_5.
 \end{eqnarray}
 $m_N$ and $m_Y$ are the masses of initial and final baryons, and $q_\mu (= p_\mu^{\prime}-p_\mu)$ is the four momentum transfer with $ Q^2 = -q^2,~Q^2 \ge 0$. 
The six form factors $f^{NY}_i(Q^2)$ and $g^{NY}_i(Q^2)$ ($i=1-3$) are determined using following assumptions about the vector and
axial vector currents in weak interactions:
\begin{enumerate}
 \item [(a)]The assumptions of T--invariance, G--invariance and SU(3) symmetry have been used to determine all the form factors $f^{NY}_i(Q^2)$ and $g^{NY}_i(Q^2)$ 
 defined in Eqs.~\ref{vec} and \ref{axi} respectively.
 
 \item[(b)]  For the determination of vector form factors we have assumed CVC which leads to $f^{NY}_3(Q^2) = 0$. 
The remaining two vector form factors $f^{NY}_1(Q^2)$ and $f^{NY}_2(Q^2)$ are determined 
in terms of the electromagnetic form factors of nucleon i.e. 
 $f_{1}^{N}(Q^{2})$ and $f_{2}^{N}(Q^{2})$ and are tabulated in Table-\ref{tab:formfac} for different processes given in Eq.~\ref{reaction}. The electromagnetic form factors of nucleon i.e. 
 $f_{1}^{N}(Q^{2})$ and $f_{2}^{N}(Q^{2})$ are in turn written in terms of Sach's electric($G_{E}^{p, n}(Q^2)$) and magnetic($G_{M}^{p, n}(Q^2)$) form factors.
 The details are given in Ref.~\cite{Alam:2014bya}.
 
There are various parameterizations for the vector form factors given in 
literature~\cite{Bradford:2006yz,Galster:1971kv,Platchkov:1989ch,Punjabi:2015bba,Bosted:1994tm,Budd:2005tm,Alberico:2008sz,Kelly:2004hm}.
We use the parameterization given by Bradford et al.~\cite{Bradford:2006yz} known as BBBA05 in all the numerical calculations presented here except in the case of $\bar\nu_\mu n \to \mu^+ \Sigma^-$, 
where sensitivity of our results to the charge form factor of neutron $G_E^n(Q^2)$ is discussed, and parameterizations of $G_E^n(Q^2)$ due to Galster et al.~\cite{Galster:1971kv,Platchkov:1989ch}
and Kelly~\cite{Punjabi:2015bba} have also been considered. 

\item[(c)]  In the axial vector sector, the form factor $g^{NY}_{2}(Q^{2})$ vanishes due to G--invariance, T--invariance and SU(3) symmetry and 
the axial vector form factor $g^{NY}_{1}(Q^{2})$ is given in terms of the axial form factor $g_A(Q^2)$ corresponding to n $\to$ p transitions. $x$ is a parameter which describes the
ratio of symmetric and antisymmetric coupling in the analysis of hyperon semileptonic decays(HSD) and is determined phenomenologically from the experimental data~\cite{Cabibbo:2003cu}. For each reaction 
considered in this work(Eq.~\ref{reaction}), the form factor $g_1^{NY}(Q^2)$ is given in Table-\ref{tab:formfac}.
A dipole parameterization for $g_A(Q^2)$ has been used 
with axial dipole mass $M_A$ i.e. 
    \begin{equation}\label{ga}
     g_{A}(Q^{2}) = g_{A}(0) \left(1 + \frac{Q^2}{M_{A}^{2}}\right)^{-2},    
    \end{equation}
with $g_A(0) = 1.2723$ determined from data on the $\beta$ decay of neutron~\cite{PDG}.  The numerical value of dipole mass $M_A$ used 
in this work is discussed in section II(f) below. 

\item[(d)]  
The pseudoscalar form factor $g_3^{NY}(Q^2)$ is obtained in terms of axial vector form factor $g_1^{NY}(Q^2)$ 
assuming PCAC and Goldberger--Treiman (GT) relation extended to strangeness sector.
We use expressions given by Marshak et al.~\cite{Marshak} and Nambu~\cite{Nambu:1960xd} where further details can be found.
Explicitly, in our numerical calculations we use the following expressions for the pseudoscalar form factor $g_3^{NY}(Q^2)$, 
  \begin{enumerate}
       \item[(i)] Marshak et al.~\cite{Marshak}:
  \begin{equation}\label{g3_marshak}
   g_3^{NY}(Q^2)=\frac{(m_N+m_Y)^2}{Q^2}\left(\frac{g_1^{NY}(Q^2)(m_K^2 + Q^2) - m_K^2 g_1^{NY}(0)}{m_K^2 + Q^2}\right),
  \end{equation}
   \item[(ii)] Nambu~\cite{Nambu:1960xd}:
   \begin{equation}\label{g3_kaon_pole}
    g_3^{NY}(Q^2)=\frac{(m_N+m_Y)^2 }{(m_K^2 +Q^2)}g_1^{NY}(Q^2),
   \end{equation}
  \end{enumerate}
  with $m_K$ being  mass of kaon and $g_1^{NY}(Q^2)$ for different $NY$ transitions is given in terms of $g_A(Q^2)$ defined in Eq.~\ref{ga}.

 \item[(e)] 
 We see from Table-\ref{tab:formfac} that SU(3) symmetry predicts a simple relation between 
 the vector and axial vector form factors for reactions 
 $\bar \nu_\mu p \to \mu^+ \Sigma^0$ and  $\bar \nu_\mu n \to \mu^+ \Sigma^-$, 
 which implies that 
 \begin{equation}
 \left[ \frac{d \sigma}{dQ^2} \right]_{ \bar \nu_\mu p \to \mu^+ \Sigma^0 }   =
\frac12 \left[ \frac{d \sigma}{dQ^2} \right]_{ \bar \nu_\mu n \to \mu^+ \Sigma^- }
 \end{equation}
and 
\begin{equation}
 \left[ P_{L,P} \right]_{ \bar \nu_\mu p \to \mu^+ \Sigma^0 } = \left[ P_{L,P} \right]_{ \bar \nu_\mu n \to \mu^+ \Sigma^- }.
\end{equation}
It should be emphasized that these relations and other implications of SU(3) symmetry and G--invariance can be tested in 
strangeness sector with the availability of precise data on weak hyperon 
production induced by antineutrinos.

\item[(f)]  
The numerical value of 
axial dipole mass($M_A$) to be used in the calculations of neutrino--nucleus cross section is a subject of intense 
discussion in the neutrino physics community and 
a wide range of $M_A$ has been recently discussed in literature~\cite{Alvarez-Ruso:2014bla,Morfin:2012kn,Gallagher:2011zza}. 
The old data available on (anti)neutrino 
scattering on hydrogen and deuterium targets~\cite{Miller:1982qi,Baker:1981su,Kitagaki:1983px}
reanalyzed by Bodek et al.~\cite{Bodek:2014pka} gives a value of $M_A=1.014\pm0.014$ GeV, while a recent analysis 
of the same data by
Meyer et al.~\cite{Meyer:2016oeg} gives a value in the range of 1.02--1.17 
GeV depending upon which data of ANL~\cite{Miller:1982qi}, BNL~\cite{Baker:1981su} 
and FNAL~\cite{Kitagaki:1983px} experiments are considered. Sometimes back, all
the world data on quasielastic (anti)neutrino
scattering from nuclear targets were analyzed by Bernard et al.~\cite{Bernard:2001rs} to yield $M_A=1.026\pm0.021$ GeV.
  
In recent years, high statistics data on quasielastic neutrino--nucleus scattering have been obtained and analyzed from neutrino and antineutrino
scattering on nuclear targets both at low and 
intermediate energies. The
data from NOMAD~\cite{nomad}, MINER$\nu$A~\cite{Fields:2013zhk}  favor a lower value of $M_A$ around 1.03 GeV, while the  
data from MiniBooNE~\cite{Mini:Aguilar},
MINOS~\cite{MINOS:Adamson}, K2K~\cite{Gran:2006jn}, T2K~\cite{Abe:2015ibe} 
and SciBooNE~\cite{Sci:Nakajima,Cheng:2012yy} favor a higher
 value of $M_A$ which lies in the range of 1.2--1.35 GeV. It is argued that at lower energies corresponding to these experiments, the (anti)neutrino quasielastic scattering from nuclear targets like
 $^{12}$C and $^{16}$O are substantially affected by the nuclear medium effects arising due to meson exchange currents(MEC), multinucleon correlations
 leading to 2p-2h and higher excitations in the nuclear medium. If these effects are adequately 
 taken into account, the low energy data can also be explained by the lower value of $M_A$ 
 around 1.03 GeV~~\cite{Alvarez-Ruso:2014bla,Morfin:2012kn,Gallagher:2011zza}. Recently, an analysis of the 
 MiniBooNE~\cite{Mini:Aguilar} and MINER$\nu$A~\cite{Fields:2013zhk} data has been done
 by Wilkinson et al.~\cite{Wilkinson:2016wmz} which concludes that these two experimental results can be explained with the
 inclusion of nuclear medium effects using a value of $M_A$ lying between 1.07--1.33 GeV.
  Furthermore, in a recent study Ankowski et al.~\cite{Ankowski:2016bji}, have analyzed experimental data from accelerator neutrinos 
  on neutrino induced reaction cross section on several nuclear targets by considering a relativistic spectral function with 2p-2h effects and found that with $M_A ~\sim$ 1.2 GeV, the data 
  on differential scattering cross section can be well explained.
  More recently the data on quasielastic cross section from MiniBooNE and MINER$\nu$A have been analyzed by Stowell et al.~\cite{Stowell:2016exm} using 
  NEUT and NuWro CCQE+2p2h
  models and it has been inferred that 
  $M_A~\sim$ 1.15 GeV can explain these experimental data.

 Keeping in view this scenario regarding the numerical values of $M_A$ to be needed to explain the quasielastic cross sections in $\Delta S=0$ (anti)neutrino--nucleus 
 scattering, we have varied
 the value of $M_A$ between 1.026--1.2 GeV in this paper to study the  $|\Delta S|=1$ quasielastic antineutrino reactions on nuclear targets. A priori, there is no reason to assume the same 
 value of $M_A$ for antineutrino quasielastic reactions in $\Delta S$ = 0 and $|\Delta S|$ = 1 sectors as argued by Gaillard and Sauvage~\cite{Gaillard:1984ny} and supported by Cabibbo 
 et al.~\cite{Cabibbo:2003cu}. However,
 this range of $M_A$ also accommodates the suggestion of 
 Gaillard and Sauvage~\cite{Gaillard:1984ny} that the value of $M_A$ to be used in $|\Delta S|=1$ quasielastic reactions should be rescaled upwards by
 a factor $\frac{m^*_K}{m_\rho}$($m^*_K$ and $m_\rho$ be 
the masses of $K^*$ and $\rho$ mesons) over the $M_A$ used in $\Delta S=0$ reactions if effects of minimal SU(3) breaking are to be simulated by taking realistic hyperons and other masses 
in the theory of HSD.

\begin{table}[h!]
 \begin{center}
\begin{adjustbox}{max width=\textwidth}  
\begin{tabular}{|c|c|c|c|c|}  \hline 
                                       &       $f_1^{NY}(Q^2)$               & $f_2^{NY}(Q^2)$      & $g_1^{NY}(Q^2)$   \\ \hline \hline

$\bar \nu_\mu p \rightarrow \mu^+ \Lambda$& $ -\sqrt{\frac{3}{2}}~f_1^p(Q^2)$& $-\sqrt{\frac{3}{2}}~f_2^p(Q^2)$ & $-\frac{1}{\sqrt{6}}(1+2x) g_A(Q^2)$\\ \hline \hline

$\bar \nu_\mu n \rightarrow \mu^+\Sigma^-$& $-\left[f_1^p(Q^2) + 2 f_1^n(Q^2) \right]$ &$-\left[f_2^p(Q^2) + 2 f_2^n(Q^2) \right]$& $(1-2x)g_A(Q^2)$ \\ \hline \hline
$\bar \nu_\mu p \rightarrow \mu^+\Sigma^0$& $-\frac{1}{\sqrt2}\left[f_1^p(Q^2) + 2 f_1^n(Q^2) \right]$ &$-\frac{1}{\sqrt2}\left[f_2^p(Q^2) + 2 f_2^n(Q^2) \right]$&
$\frac{1}{\sqrt2}(1-2x)g_A(Q^2)$ \\ \hline \hline
                           
  \end{tabular}
  \end{adjustbox}
\end{center}
\caption{Vector and axial vector from factors for $\bar{\nu}_\mu(k) + N(p)\rightarrow \mu^+(k^\prime) + Y(p^\prime)$ processes.}
 \label{tab:formfac}
\end{table}

  \end{enumerate}
\subsection{Cross section}
The differential cross section corresponding to the processes given  in Eq.~\ref{reaction} may be written as
 \begin{equation}
 \label{crosv.eq}
 d\sigma=\frac{1}{(2\pi)^2}\frac{1}{4E_{\bar \nu_\mu} m_N}\delta^4(k+p-k^\prime-p^\prime)
 \frac{d^3k^\prime}{2E_{k^\prime}}\frac{d^3p^\prime}{2E_{p^\prime}}\sum \overline{\sum} |{\cal{M}}|^2,
 \end{equation}
where ${\cal{M}}$ is the transition matrix element, square of which may be 
written in terms of hadronic and leptonic tensors as 
\begin{equation}
 |{\cal{M}}|^2 = \frac{G_F^2 \sin^2\theta_c}{2} \cal{J}^{\alpha \beta} \cal{L}_{\alpha \beta}. 
\end{equation}
The hadronic and leptonic tensors are given by
\begin{eqnarray}\label{JL}
\cal{J}^{\alpha \beta} &=& \mathrm{Tr}\left[\Lambda(p')J^{\alpha}
 \Lambda(p)\tilde{J}^{\beta} \right] \nonumber \\
 \cal{L}_{\alpha \beta} &=& \mathrm{Tr}\left[\gamma_{\alpha}(1+\gamma_{5})k\!\!\!/
\gamma_{\beta}(1+\gamma_{5})(k'\!\!\!\!/+m_\mu)\right],
\end{eqnarray} 
with $\tilde{J}_{\beta} =\gamma^0 J^{\dagger}_{\beta} \gamma^0$ and $\Lambda(p)=p\!\!\!/+m_N$. 
Using the above definitions, the $Q^2$ distribution is written as
\begin{equation}\label{dsig}
 \frac{d\sigma}{dQ^2}=\frac{G_F^2 \sin^2\theta_c}{8 \pi m_N E_{\bar \nu_{_\mu}}^2} {\cal N}(Q^2,E_{\bar \nu_{_\mu}}),
\end{equation}
where the expression of ${\cal N}(Q^2,E_{\bar \nu_{_\mu}})$ is given in the appendix.

When the reactions shown in Eq.~\ref{reaction} take place on nucleons which are bound in the
nucleus, the neutrons and protons are not free and their momenta $p_{n,p}(r)$ at r are constrained to satisfy the Pauli principle, 
i.e., ${p_{n,p}(r)}<{p_{F_{n,p}}(r)}$, where $p_{F_n}(r)$ and $p_{F_p}(r)$ are the
local Fermi momenta of neutrons and protons at the interaction point in the nucleus and are given by 
$p_{F_n}(r)=\left[3\pi^2\rho_n(r)\right]^\frac{1}{3}$ and $p_{F_p}(r)=\left[3\pi^2\rho_p(r)\right]^\frac{1}{3}$, $\rho_n(r)$
and $\rho_p(r)$ are the neutron and proton nuclear densities given by $\rho_{n}(r)=\frac{(A-Z)}{A}\rho(r)$ and $\rho_{p}(r)=\frac{Z}{A}\rho(r)$,  $\rho(r)$ is the nuclear density which is determined
 from electron-nucleus scattering experiments.

The differential scattering cross section for the scattering of antineutrinos 
from nucleons in the nucleus is then given as 
\begin{equation}\label{diffnuc}
\left[\frac{d^2\sigma}{dE_ld\Omega_l}\right]_{\bar{\nu}_{_\mu}A}=2{\int_{r_{min}}^{r_{max}} d^3r \int_{0}^{p_{F_N}(r)} \frac{d^3p}{{(2\pi)}^3}n_N(p,r)
\left[\frac{d^2\sigma}{dE_ld\Omega_l}\right]_{\bar{\nu}_{_\mu}N}}
\end{equation}
where $n_N(p,r)$ is local occupation number of the initial
nucleon of momentum $p$ at a radius $r$ in the nucleus, which is 1 for $p < p_{F_N}(r)$ and 0 otherwise, and $n_N(p,r)$ is related to the density as:
\begin{eqnarray}
\rho = \frac NV =2 \int \frac{d^3p}{{(2\pi)}^3}n_N(p,r).
\end{eqnarray}

In the next section, we discuss briefly the construction of 
polarization vector for the final hyperon. 
\subsection{Polarization of hyperons}
Using the covariant density matrix formalism, polarization 4-vector($\xi^\tau$) of the final hyperon produced
in reaction~(\ref{hyp-rec}) is written as~\cite{Bilekny}:
\begin{equation}\label{polar}
\xi^{\tau}=\frac{\mathrm{Tr}[\gamma^{\tau}\gamma_{5}~\rho_{f}(p^\prime)]}
{\mathrm{Tr}[\rho_{f}(p^\prime)]},
\end{equation}
where the final spin density matrix $\rho_f(p^\prime)$ is given by 
\begin{equation}\label{polar1}
 \rho_{f}(p^\prime)= {\cal L}^{\alpha \beta}  \Lambda(p')J_{\alpha} \Lambda(p)\tilde{J}_{\beta} 
\Lambda(p').
\end{equation} 
Using the following relations~\cite{Bilenky:2013fra,Bilenky:2013iua}
\begin{equation}\label{polar3}
\Lambda(p')\gamma^{\tau}\gamma_{5}\Lambda(p')=2m_Y\left(g^{\tau\sigma}-
\frac{p'^{\tau}p'^{\sigma}}{m_Y^{2}}\right)\Lambda(p')\gamma_{\sigma}
\gamma_{5}
\end{equation}
and
\begin{equation}
 \Lambda(p^\prime)\Lambda(p^\prime) = 2m_Y \Lambda(p^\prime),
\end{equation}
 $\xi^\t$ defined in Eq.~\ref{polar} may be rewritten as:
\begin{equation}\label{polar4}
\xi^{\tau}=\left( g^{\tau\sigma}-\frac{p'^{\tau}p'^{\sigma}}{m_Y^2}\right)
\frac{  {\cal L}^{\alpha \beta}  \mathrm{Tr}
\left[\gamma_{\sigma}\gamma_{5}\Lambda(p')J_{\alpha} \Lambda(p)\tilde{J}_{\beta} \right]}
{ {\cal L}^{\alpha \beta} \mathrm{Tr}\left[\Lambda(p')J_{\alpha} \Lambda(p)\tilde{J}_{\beta} \right]}.
\end{equation}
Note that in Eq.~\ref{polar4}, $\xi^\t$ is manifestly orthogonal to $p^{\prime \t}$ i.e. $p^\prime \cdot \xi=0$. Moreover, the denominator
is directly related to the differential cross section given in Eq.~\ref{dsig}.

With ${\cal J}^{\a \b}$ and ${\cal L}_{\a \b}$ given in Eq.~\ref{JL}, an expression for $\xi^\t$ is obtained. In the lab frame where the initial nucleon 
is at rest, the polarization vector $\vec \xi$ is calculated to be 
\begin{equation}\label{3pol}
\frac{d\sigma}{dQ^2} \vec \xi =\frac{G_F^2 \sin^2 \theta_c }{8 \pi\; m_N m_Y E^2_{\bar \nu_\mu}}\left[(\vec k + \vec k^{\prime})m_Y {\cal A}(Q^2,E_{\bar \nu_{_\mu}}) + (\vec k -
\vec k^{\prime}) {\cal B}(Q^2,E_{\bar \nu_{_\mu}}) \right], 
\end{equation}
where the expressions of ${\cal A}(Q^2,E_{\bar \nu_{_\mu}})$ and ${\cal B}(Q^2,E_{\bar \nu_{_\mu}})$ are given in the
appendix.

From Eq.~\ref{3pol}, it follows that the polarization lies in scattering plane defined by $\vec k$ and $\vec k^\prime$, and there is no component of
polarization in a direction orthogonal to the scattering 
plane. This is a consequence of T--invariance which makes the transverse polarization in a direction perpendicular to the reaction plane to
vanish~\cite{Pais:1971er,LlewellynSmith:1971uhs}. We now expand the polarization vector $\vec \xi$ along two orthogonal directions, $\vec e_L$ and $\vec e_P$ in the reaction plane
corresponding to parallel and perpendicular 
directions to the momentum of hyperon\footnote{It should be noted that our $\vec e_P$ is defined as in Bilenky and Christova~\cite{Bilenky:2013fra}
and is opposite to the  sign used by Erriquez et al.~\cite{Erriquez:1977tr}.} i.e.
\begin{equation}\label{vectors}
\vec{e}_{L}=\frac{\vec{p'}}{|\vec{p'}|}=\frac{\vec{q}}{|\vec{q}|},\qquad
\vec{e}_{P}=\vec{e}_{L}\times \vec{e_T},\qquad  \vec{e_T}=\frac{\vec{k}\times\vec{k'}}{|\vec{k}\times\vec{k'}|},
\end{equation}
and write
 \begin{equation}\label{polarLab}
\vec{\xi}=\xi_{P}\vec{e}_{P}+\xi_{L}\vec{e}_{L},
\end{equation}
such that the longitudinal and perpendicular components of polarization vector($\vec \xi$) in the lab frame are given by
\begin{equation}\label{PL}
 \xi_L(Q^2)=\vec \xi \cdot \vec e_L,\qquad \xi_P(Q^2)=\vec \xi \cdot \vec e_P.
\end{equation}
From Eq.~\ref{PL}, the longitudinal and perpendicular components of polarization vector $P_L(Q^2)$ and $P_P(Q^2)$ defined in the rest frame of recoil nucleon are given by ~\cite{Bilenky:2013fra}:
\begin{equation}\label{PL1}
 P_L(Q^2)=\frac{m_Y}{E_{p^\prime}} \xi_L(Q^2), \qquad P_P(Q^2)=\xi_P(Q^2),
\end{equation}
where $\frac{m_Y}{E_{p^\prime}}$ is the Lorentz boost factor along $\vec p^\prime$.
With the help of Eqs.~\ref{3pol}, \ref{vectors}, \ref{PL} and \ref{PL1}, the longitudinal component $P_L(Q^2)$ is calculated to be
\begin{equation}\label{sl}
\frac{d\sigma}{dQ^2} P_L(Q^2)= \frac{G_F^2 \sin^2\theta_c}{8\pi|\vec q| E_{p^{\prime}} m_N\;E^2_{\bar \nu_{_\mu}}} \left[\left(E_{\bar \nu_{_\mu}}^2 - E_\mu^2 + m_\mu^2 \right)m_Y
  {\cal A}(Q^2,E_{\bar \nu_{_\mu}}) + |\vec q|^2 {\cal B}(Q^2,E_{\bar \nu_{_\mu}})\right],
\end{equation}
where in the lab frame $E_{p^{\prime}} = \sqrt{|{\vec q}^2|+m_Y^2}$. 
Similarly, the perpendicular component $P_P(Q^2)$  of the polarization 3-vector is given as
\begin{equation}\label{st}
\frac{d\sigma}{dQ^2} P_P(Q^2) = -\frac{G_F^2 \sin^2\theta_c}{4\pi} \frac{|\vec k^{\prime }|}{|\vec q|}\frac{{\cal A}(Q^2,E_{\bar \nu_{_\mu}}) \sin\theta}{m_N E_{\bar \nu_{_\mu}}},
\end{equation}
where $\theta$ is the scattering angle in the lab frame.

Inside the nucleus, target nucleon is not at rest but moves with Fermi momentum, i.e. $\vec{p} \ne 0$. Because of this the polarization components of the final hyperon get modified to:\\
\begin{equation}\label{pol_nuc}
\left[P_{L,P}(Q^2)\right]_{\bar{\nu}_{_\mu}A}=2{\int d^3r \int \frac{d^3p}{{(2\pi)}^3}n_N(p,r)
\left[P_{L,P}(Q^2,\vec{p})\right]_{\bar{\nu}_{_\mu}N}},
\end{equation}

  with longitudinal component:\\ 
 
 \begin{eqnarray}\label{PL_nuc}
                                  P_L(Q^2,\vec{p}) = \frac{m_Y}{E_{p^\prime}}\frac{G_F^2 \sin^2\theta_c}{2} && \frac{1}{|{\cal M}|^2} \frac{1}{|\vec{p}+\vec{q}|}
 \left[ \alpha(Q^2,\vec{p}) \left( \vec{k}\cdot\vec{p}+E^2_{\bar\nu_{_\mu}}-\vec{k}\cdot\vec{k^\prime}\right)\right. \nonumber \\
 &&\left.  + ~\beta(Q^2,\vec{p})\left(\vec{k^\prime}\cdot\vec{p}+\vec{k}\cdot\vec{k^\prime}-|\vec{k^\prime}|^2\right) +\eta(Q^2,\vec{p}) \left(|\vec{p}|^2+\vec{p}\cdot\vec{q}\right)\right],
                                 \end{eqnarray}

 and perpendicular component,\\\begin{eqnarray}\label{Pp_nuc}
                                                    P_P (Q^2,\vec{p})&=&\frac{G_F^2 \sin^2\theta_c}{2}\frac{1}{|{\cal M}|^2}\frac{1}{|\vec{p}+\vec{q}||\vec{k}||\vec{k^\prime}|\sin \theta}
 \left[ \left(\vec{k^\prime}\cdot\vec{p}+\vec{k}\cdot\vec{k^\prime}-|\vec{k^\prime}|^2\right)\{\alpha(Q^2,\vec{p}) E^2_{\bar \nu_{_\mu}}
 +\beta(Q^2,\vec{p}) \vec{k}\cdot\vec{k^\prime}\right. \nonumber\\
 &+& \left.\eta(Q^2,\vec{p}) \vec{k}\cdot\vec{p}\} - \left(\vec{k}\cdot\vec{p}+E^2_{\bar\nu_{_\mu}}-\vec{k}\cdot\vec{k^\prime}\right)\{\alpha(Q^2,\vec{p}) \vec{k}\cdot\vec{k^\prime}
 +\beta(Q^2,\vec{p}) |\vec{k^\prime}|^2 + 
 \eta(Q^2,\vec{p}) \vec{k^\prime}\cdot\vec{p}\}\right].
                                                    \end{eqnarray}

The expressions of $\alpha(Q^2,\vec{p})$, $\beta(Q^2,\vec{p})$ and $\eta(Q^2,\vec{p})$ are given in the appendix.
\section{Results and Discussion}\label{Results and Discussion}
\subsection{Differential cross section $\frac{d\sigma}{dQ^2}$ and polarization components $P_L(Q^2)$ and $P_P(Q^2)$ for nucleon target}
\begin{figure}
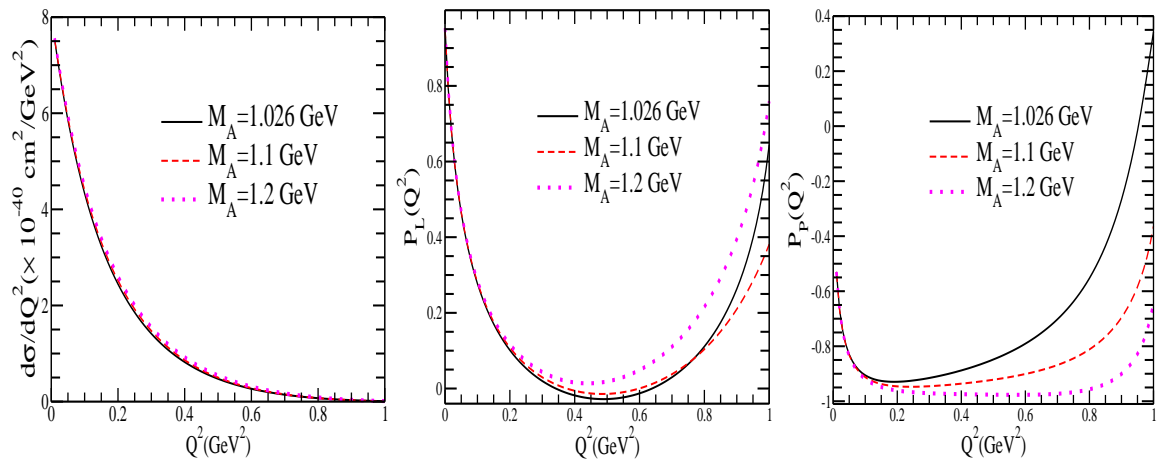

   \includegraphics[height=6cm,width=5cm]{q2_dstbn_Ma_hyp.eps}
   \includegraphics[height=6cm,width=5cm]{sL_dstbn_Ma_lam_enu1.eps}
   \includegraphics[height=6cm,width=5cm]{sT_dstbn_Ma_hyp.eps}
   \caption{ $\frac{d\sigma}{dQ^2}$, $P_L(Q^2)$ and $P_P(Q^2)$ vs $Q^2$ for the process $\bar \nu_\mu p \to \mu^+ \Lambda$ at $E_{\bar \nu_{_\mu}}$ = 1 GeV 
   for different values of $M_A$ used in $g_1^{p\Lambda}(Q^2)$ 
   viz. 1.026 GeV(solid line), 1.1 GeV(dashed) and
  1.2 GeV(dotted line)  with  $m_\mu=0 $. $f_1^{p\Lambda}(Q^2),~f_2^{p\Lambda}(Q^2)~ \text{and} ~g_1^{p\Lambda}(Q^2)$ from Table-\ref{tab:formfac} 
  and BBBA05 parameterization for nucleon form factor are used.}\label{Ma_dstbn}
\end{figure}
\begin{figure}
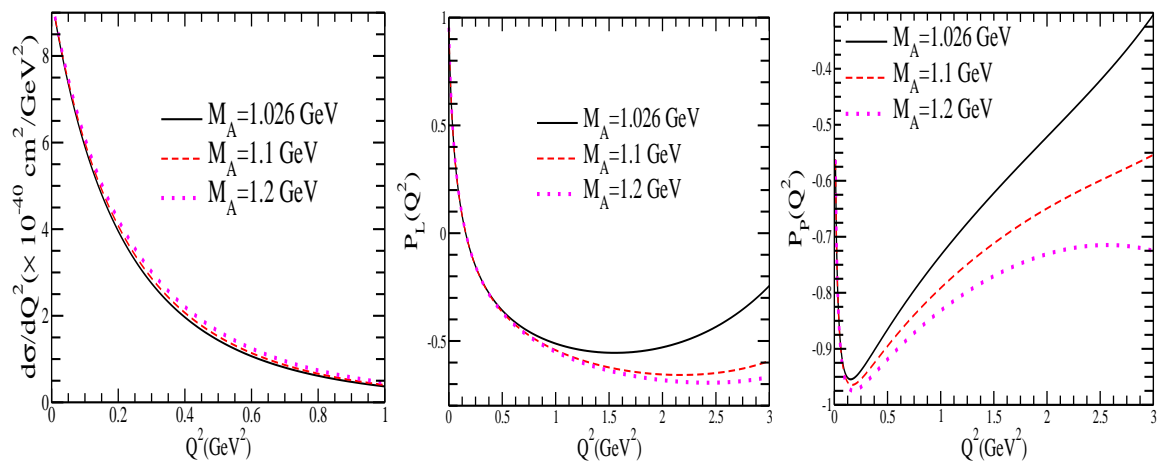

   \includegraphics[height=6cm,width=5cm]{q2_dstbn_Ma_hyp_enu3.eps}
   \includegraphics[height=6cm,width=5cm]{sL_dstbn_Ma_lam_enu3.eps}
   \includegraphics[height=6cm,width=5cm]{sT_dstbn_Ma_hyp_enu3.eps}
   \caption{ $\frac{d\sigma}{dQ^2}$, $P_L(Q^2)$ and $P_P(Q^2)$ vs $Q^2$ for the process $\bar \nu_\mu p \to \mu^+ \Lambda$ at $E_{\bar \nu_{_\mu}}$ = 3 GeV for different values of $M_A$
   in $g_1^{p\Lambda}(Q^2)$. Lines and points have the same meaning as in Fig.~\ref{Ma_dstbn}.}\label{Ma_dstbn_enu3}
\end{figure}
We have used Eqs.~\ref{dsig}, \ref{sl} and \ref{st} to numerically evaluate the differential cross section $\frac{d\sigma}{dQ^2}$, and 
 longitudinal $P_L(Q^2)$ and perpendicular $P_P(Q^2)$ components of the polarization of hyperons in the quasielastic antineutrino reactions given in Eq.~\ref{reaction}.
For the vector and axial vector form factors we have used the expressions of $f_i^{NY}(Q^2)(i=1,2)$ and $g_1^{NY}(Q^2)$ given in Table-\ref{tab:formfac}
along with the pseudoscalar form factor $g_3^{NY}(Q^2)$ given in Eqs.~\ref{g3_marshak} and \ref{g3_kaon_pole}. The $Q^2$
dependence of the nucleon form factors $f_{1,2}^{p,n}$ is taken from the
parameterization of BBBA05~\cite{Bradford:2006yz}. A dipole 
parameterization for the axial vector form factor $g_A(Q^2)$ 
given in Eq.~\ref{ga} has been used for $g_{1,3}^{NY}(Q^2)$ with $g_A(0) = 1.2723$~\cite{PDG}, $x=0.364$~\cite{Cabibbo:2003cu}  and axial dipole mass $M_A~=$ 1.026 GeV, 1.1 GeV 
and 1.2 GeV as mentioned in each figure. 

In Fig.~\ref{Ma_dstbn}, we present the results of $\frac{d\sigma}{dQ^2}$, $P_L(Q^2)$ and $P_P(Q^2)$ for the reaction
$\bar\nu_\mu p \to \mu^+ \Lambda$ at $E_{\bar \nu_{_\mu}}=1$ GeV and in Fig.~\ref{Ma_dstbn_enu3} at $E_{\bar \nu_{_\mu}}=3$ GeV.
We see that while there is very little sensitivity of $\frac{d\sigma}{dQ^2}$ to the variation of $M_A$, the components of 
polarization $P_L(Q^2)$ and $P_P(Q^2)$ are quite 
sensitive to the value of $M_A$ specially in the region $Q^2>0.4$ GeV$^2$.
It should, therefore, be possible to independently determine the value of $M_A$ from the
polarization measurements. However, the present
available data on the total cross section for the single hyperon production are
consistent with $M_A=1.026$ GeV~\cite{Alam:2014bya}.
At higher values of $Q^2$, the sensitivity of $P_L(Q^2)$ and $P_P(Q^2)$ to $M_A$ increases, but quantitatively, the cross section $\frac{d\sigma}{dQ^2}$ decreases, making the
number of events quite small and the measurement of polarization observables becomes difficult.
We have also studied the sensitivity of our results for $\frac{d\sigma}{dQ^2}$, $P_L(Q^2)$ and $P_P(Q^2)$ 
to various other parameterizations of $Q^2$ dependence of the nucleon form factors $f_{1,2}^{p,n}(Q^2)$
available in literature~\cite{Bradford:2006yz,Galster:1971kv,Platchkov:1989ch,Punjabi:2015bba,Bosted:1994tm,Budd:2005tm,Alberico:2008sz,Kelly:2004hm}. 
It is found that at $E_{\bar \nu_{_\mu}} =1$ GeV, the results for $\frac{d\sigma}{dQ^2}$, $P_L(Q^2)$ and
$P_P(Q^2)$ are not very sensitive to the choice of other parameterizations of vector form factors in the case of  
$\bar \nu_\mu p \to \mu^+ \Lambda$ and are not shown in these figures.
\begin{figure}
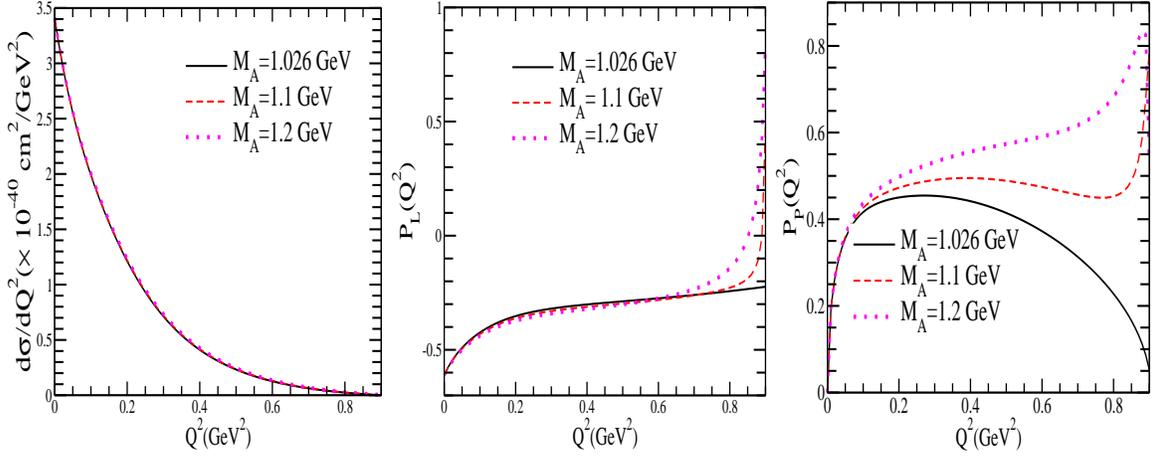

   \includegraphics[height=6cm,width=5cm]{q2_dstbn_Ma_sig_minus_enu1.eps}
   \includegraphics[height=6cm,width=5cm]{sL_dstbn_Ma_sig_minus_enu1.eps}
   \includegraphics[height=6cm,width=5cm]{sT_dstbn_Ma_sig_minus_enu1.eps}
   \caption{ $\frac{d\sigma}{dQ^2}$, $P_L(Q^2)$ and $P_P(Q^2)$ vs $Q^2$ for the process $\bar \nu_\mu n \to \mu^+ 
   \Sigma^-$ at $E_{\bar \nu_{_\mu}}$= 1 GeV for different values of $M_A$ in $g_1^{n \Sigma^-}(Q^2)$. 
Lines and points have the same meaning as in Fig.~\ref{Ma_dstbn}.}\label{Ma_dstbn_enu1_sig}
\end{figure}
\begin{figure}
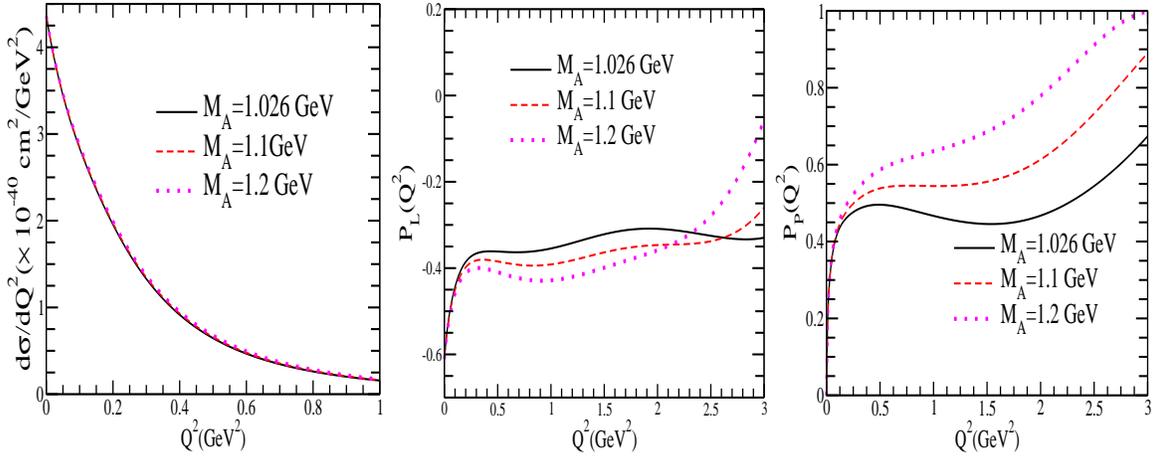

   \includegraphics[height=6cm,width=5cm]{q2_dstbn_Ma_sig_minus_enu3.eps}
   \includegraphics[height=6cm,width=5cm]{sL_dstbn_Ma_sig_minus_enu3.eps}
   \includegraphics[height=6cm,width=5cm]{sT_dstbn_Ma_sig_minus_enu3.eps}
   \caption{ $\frac{d\sigma}{dQ^2}$, $P_L(Q^2)$ and $P_P(Q^2)$ vs $Q^2$ for the process $\bar \nu_\mu n \to \mu^+ \Sigma^-$ at $E_{\bar \nu_{_\mu}}$= 3 
   GeV for different values of $M_A$ in $g_1^{n \Sigma^-}(Q^2)$. 
Lines and points have the same meaning as in Fig.~\ref{Ma_dstbn}.}\label{Ma_dstbn_enu3_sig}
\end{figure}
 
In Figs.~\ref{Ma_dstbn_enu1_sig} and \ref{Ma_dstbn_enu3_sig}, we present the results of $\frac{d\sigma}{dQ^2}$,
$P_L(Q^2)$ and $P_P(Q^2)$ for the reaction
$\bar\nu_\mu n \to \mu^+ \Sigma^-$ at $E_{\bar \nu_{_\mu}}=1$ GeV and $E_{\bar \nu_{_\mu}}=3$ GeV, respectively. The results for 
$\frac{d\sigma}{dQ^2}$, $P_L(Q^2)$ and $P_P(Q^2)$ are qualitatively similar to $\bar\nu_\mu p \to \mu^+\Lambda$ as
far as the sensitivity to $M_A$ is concerned. However, the differential cross sections
$\frac{d\sigma}{dQ^2}$ are smaller and the components of the hyperon polarization are of the same order as in reaction 
$\bar \nu_\mu p \to \mu^+ \Lambda$ but slightly higher in magnitude. We have chosen to show the results 
for $\bar\nu_\mu n \to \mu^+ \Sigma^-$ as the cross section for this process is larger by a factor of 2 
as compared to $\bar\nu_\mu p \to \mu^+ \Sigma^0$. While there is very little sensitivity of $\frac{d\sigma}{dQ^2}$,
$P_L(Q^2)$ and $P_P(Q^2)$ to the vector form factors in the case of $\bar \nu_\mu p \to \mu^+ \Lambda$, this is not
the case for $\bar \nu_\mu n \to \mu^+ \Sigma^-$. In the case of  $\bar \nu_\mu n \to \mu^+ \Sigma^-$ process, the results for differential cross section $\frac{d\sigma}{dQ^2}$ and
polarization components $P_{L}(Q^2)$ and $P_{P}(Q^2)$ are found to be sensitive to the vector form factors specially to the neutron form factors $f_{1,2}^n(Q^2)$ occurring in the expressions 
of $f^{n\Sigma^-}_{1,2}$(see Table-\ref{tab:formfac}). 
This arises mainly due to the presence of charge form factor of neutron $G_E^n(Q^2)$ in the definition of $f_{1,2}^n(Q^2)$.
 We have, therefore, studied the sensitivity of our
 results to various parameterizations of charge form factor of neutron available in literature. Some of the different parameterizations for $G_E^n(Q^2)$ being used recently in the literature are
 \cite{Bradford:2006yz,Galster:1971kv,Platchkov:1989ch,Punjabi:2015bba}:

\begin{itemize}
 \item Bradford et al.(BBBA05)~\cite{Bradford:2006yz}:
 \begin{equation}
  G_E^n(Q^2) = \frac{a_1 \tau+ a_2 \tau^2}{1+b_1 \tau+ b_2 \tau^2+ b_3 \tau^3},
 \end{equation}
with $a_1$=1.25, $a_2$=1.30, $b_1$= -9.86, $b_2$= 305.0 and $b_3$=7.54.

\item Galster et al.~\cite{Galster:1971kv}:
\begin{equation}
 G_E^n(Q^2) = -\frac{\mu_n \tau}{1+5.6\tau}G_D(Q^2),
\end{equation}
with $\mu_n$=$-$1.913, $\tau=\frac{Q^2}{4m_N^2}$, $G_D(Q^2) = \left(1+\frac{Q^2}{M_V^2}\right)^{-2}$;  $M_V=0.84$GeV.

\item Modified form of $G_E^n(Q^2)$ in Galster et al. parameterization ~\cite{Platchkov:1989ch}:
\begin{equation}
 G_E^n(Q^2) = -\frac{a \mu_n \tau}{1+b\tau}G_D(Q^2),
\end{equation}
with $a$=1.51 and $b$=8.4.

\item Modified form of $G_E^n(Q^2)$ in Kelly parameterization~\cite{Punjabi:2015bba}:
\begin{equation}
 G_E^n(Q^2) = \frac{G_M^n(Q^2)}{\mu_n} \frac{a_1\tau}{1+a_2 \sqrt{\tau}+a_3\tau},
 \end{equation}
 with $a_1$=2.6316, $a_2$=4.118 and $a_3$=0.29516.
\end{itemize}

\begin{figure}
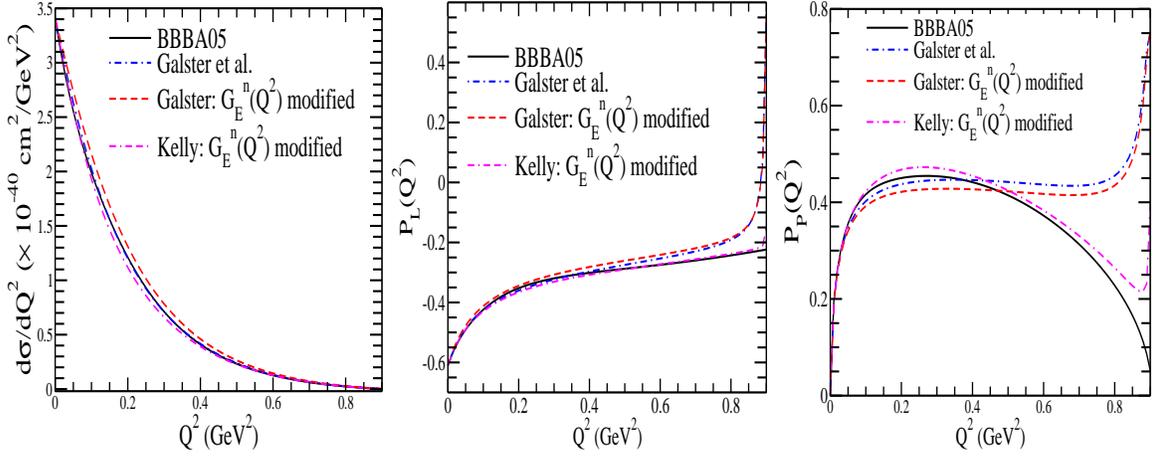

   \includegraphics[height=6cm,width=5cm]{q2_gen_sig_enu1.eps}
   \includegraphics[height=6cm,width=5cm]{sL_gen_sig_enu1.eps}
   \includegraphics[height=6cm,width=5cm]{sT_gen_sig_enu1.eps}
   \caption{$\frac{d\sigma}{dQ^2}$, $ P_L(Q^2)$ and $P_P(Q^2)$ vs $Q^2$ at $E_{\bar \nu_{_\mu}}$= 1 GeV for $\bar\nu_\mu n \to \mu^+ \Sigma^-$ process.
The results are presented with the 
  nucleon form factors using BBBA05~\cite{Bradford:2006yz}(solid line), Galster et al.~\cite{Galster:1971kv}(dashed-dotted line), modified form of $G_E^n(Q^2)$ 
  in Galster parameterization~\cite{Platchkov:1989ch}(dashed line) and modified form of $G_E^n(Q^2)$ in Kelly parameterization~\cite{Punjabi:2015bba}(double dashed-dotted line).}\label{sig_gen}
\end{figure}

\begin{figure}
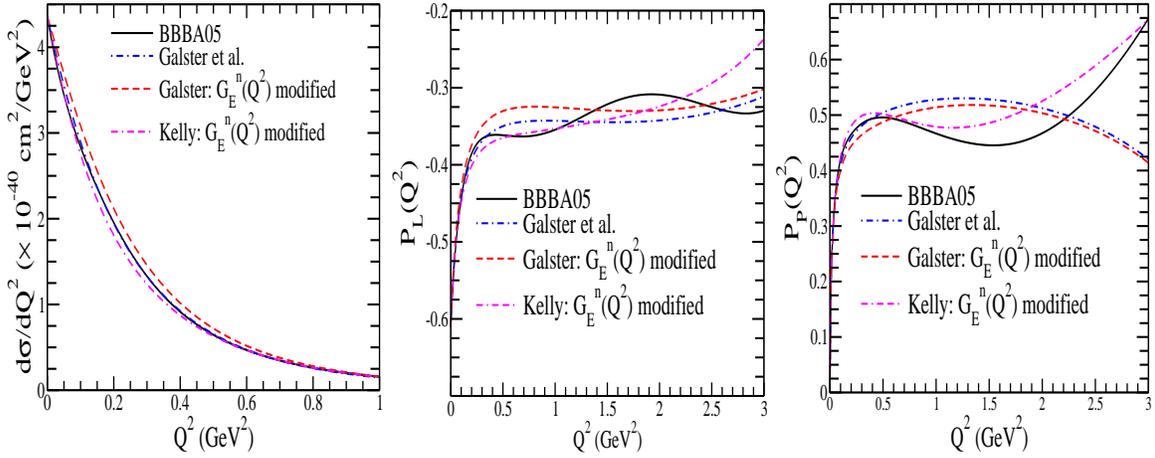

   \includegraphics[height=6cm,width=5cm]{q2_gen_sig_enu3.eps}
   \includegraphics[height=6cm,width=5cm]{sL_gen_sig_enu3.eps}
   \includegraphics[height=6cm,width=5cm]{sT_gen_sig_enu3.eps}
   \caption{$\frac{d\sigma}{dQ^2},$ $ P_L(Q^2)$ and $P_P(Q^2)$ vs $Q^2$ at $E_{\bar \nu_{_\mu}}$= 3 GeV for $\bar\nu_\mu n \to \mu^+ \Sigma^-$ process. Lines and points 
   have the same meaning as
   Fig.~\ref{sig_gen}.}\label{sig_gen_enu3}
\end{figure}

 We show in Figs.~\ref{sig_gen} ($E_{\bar \nu_{_\mu}}=1$ GeV) and \ref{sig_gen_enu3} ($E_{\bar \nu_{_\mu}}=3$ GeV), the dependence of $\frac{d\sigma}{dQ^2}$,
$P_L(Q^2)$ and $P_P(Q^2)$ on the different parameterization of $G_E^n(Q^2)$. It is seen that the polarization observables 
 are quite sensitive to the neutron charge form factor in $\bar\nu_\mu n\to \mu^+ \Sigma^-$ specially at $E_{\bar \nu_{_\mu}}=3$ GeV and it 
 should be possible to determine, in principle, the charge form factor of neutron from the observation of $P_L(Q^2)$ and $P_P(Q^2)$ using this process.


\begin{figure}
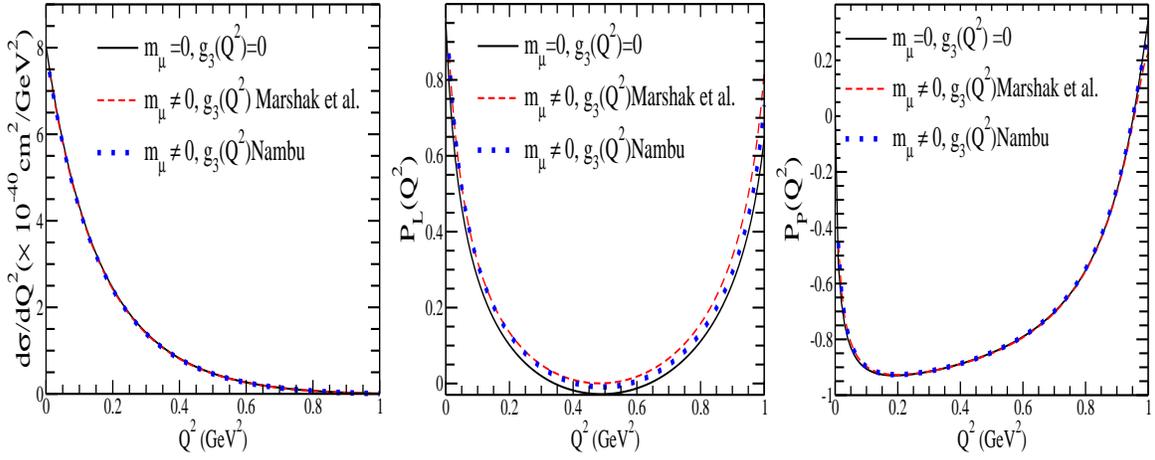

   \includegraphics[height=6cm,width=5cm]{q2_fp_vartn_hyp.eps}
   \includegraphics[height=6cm,width=5cm]{sL_dstbn_fp_vartn_lam.eps}
   \includegraphics[height=6cm,width=5cm]{sT_dstbn_fp_vartn_hyp.eps}
   \caption{ $\frac{d\sigma}{dQ^2}$, $P_L(Q^2)$ and $P_P(Q^2)$ vs $Q^2$ ($M_A=$ 1.026 GeV)
   for the process $\bar \nu_\mu p \to \mu^+ \Lambda$ at $E_{\bar \nu_\mu}=1$ GeV
   using $f_1^{p\Lambda}(Q^2),~f_2^{p\Lambda}(Q^2),~g_1^{p\Lambda}(Q^2)$ 
   from Table-\ref{tab:formfac} and BBBA05~\cite{Bradford:2006yz} parameterization for the nucleon form factors,
    with $m_\mu = 0$ and $g_3^{p\Lambda} = 0$(solid line), 
$m_\mu \ne 0$ and $g_3^{p\Lambda} \ne 0$ from Marshak et al.~\cite{Marshak} given in Eq.~\ref{g3_marshak}(dashed line) and 
$m_\mu \ne 0$ and $g_3^{p\Lambda} \ne 0$ from Nambu~\cite{Nambu:1960xd} given in Eq.~\ref{g3_kaon_pole}(dotted line).}\label{fp_vartn}

\end{figure}

\begin{figure}
   \includegraphics[height=6cm,width=5cm]{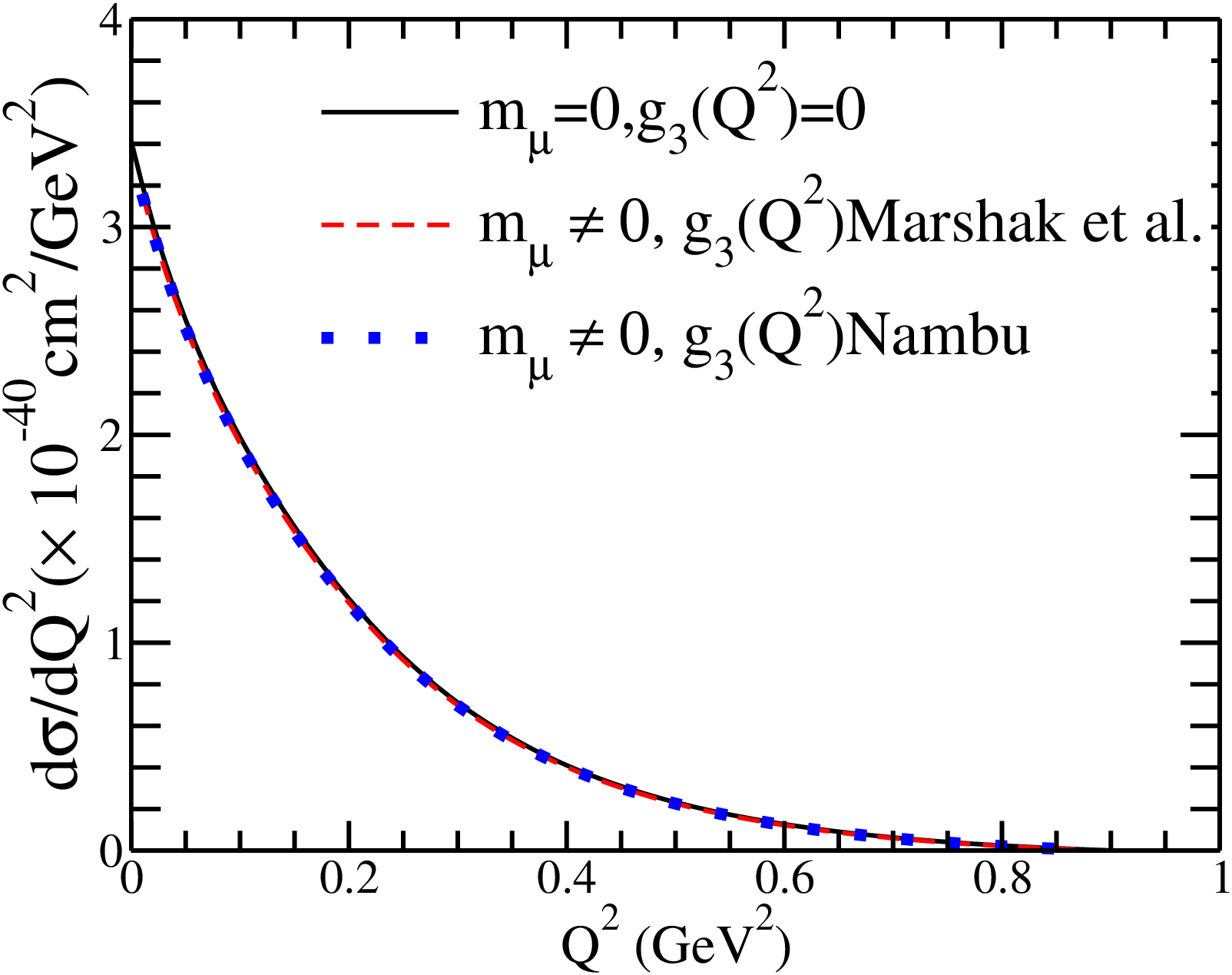}
   \includegraphics[height=6cm,width=5cm]{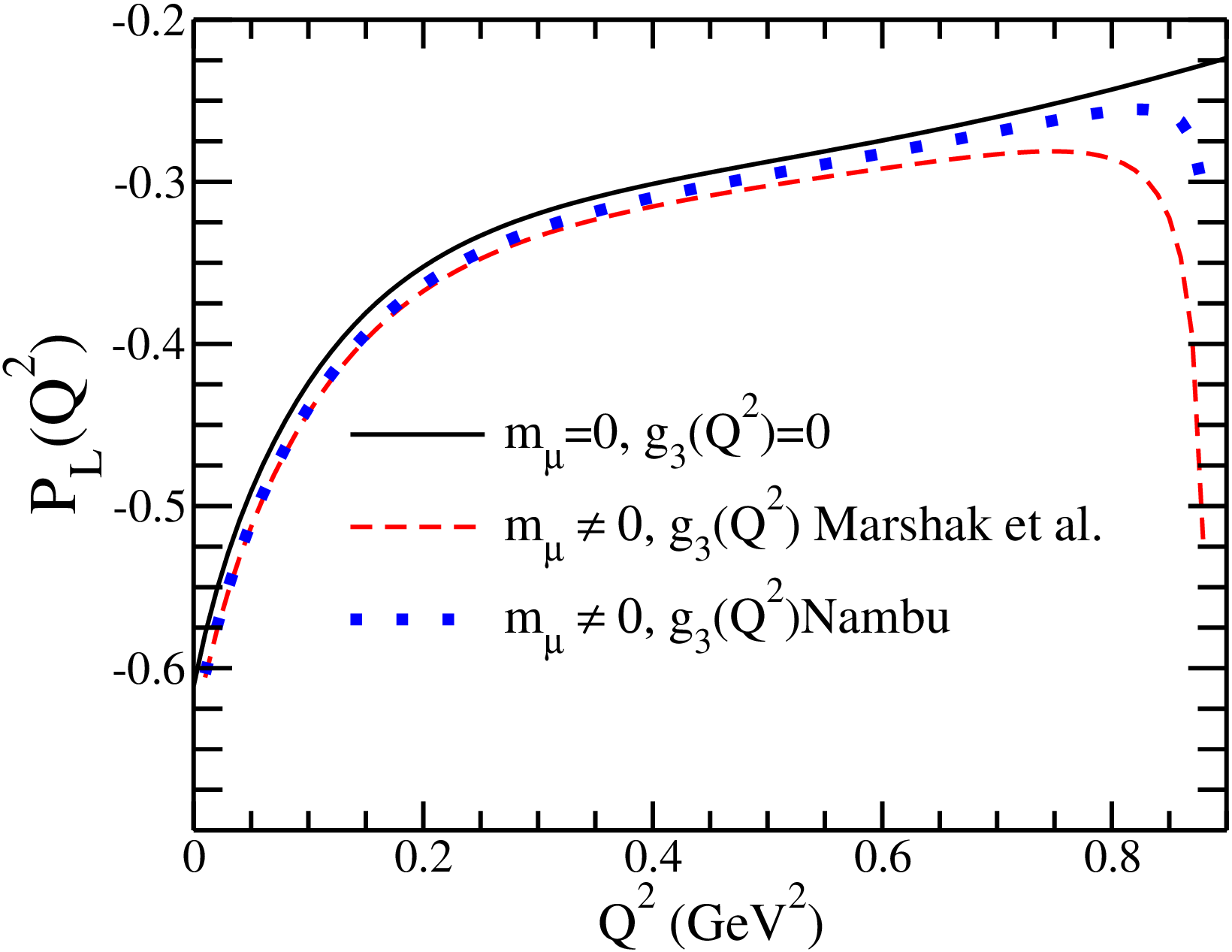}
   \includegraphics[height=6cm,width=5cm]{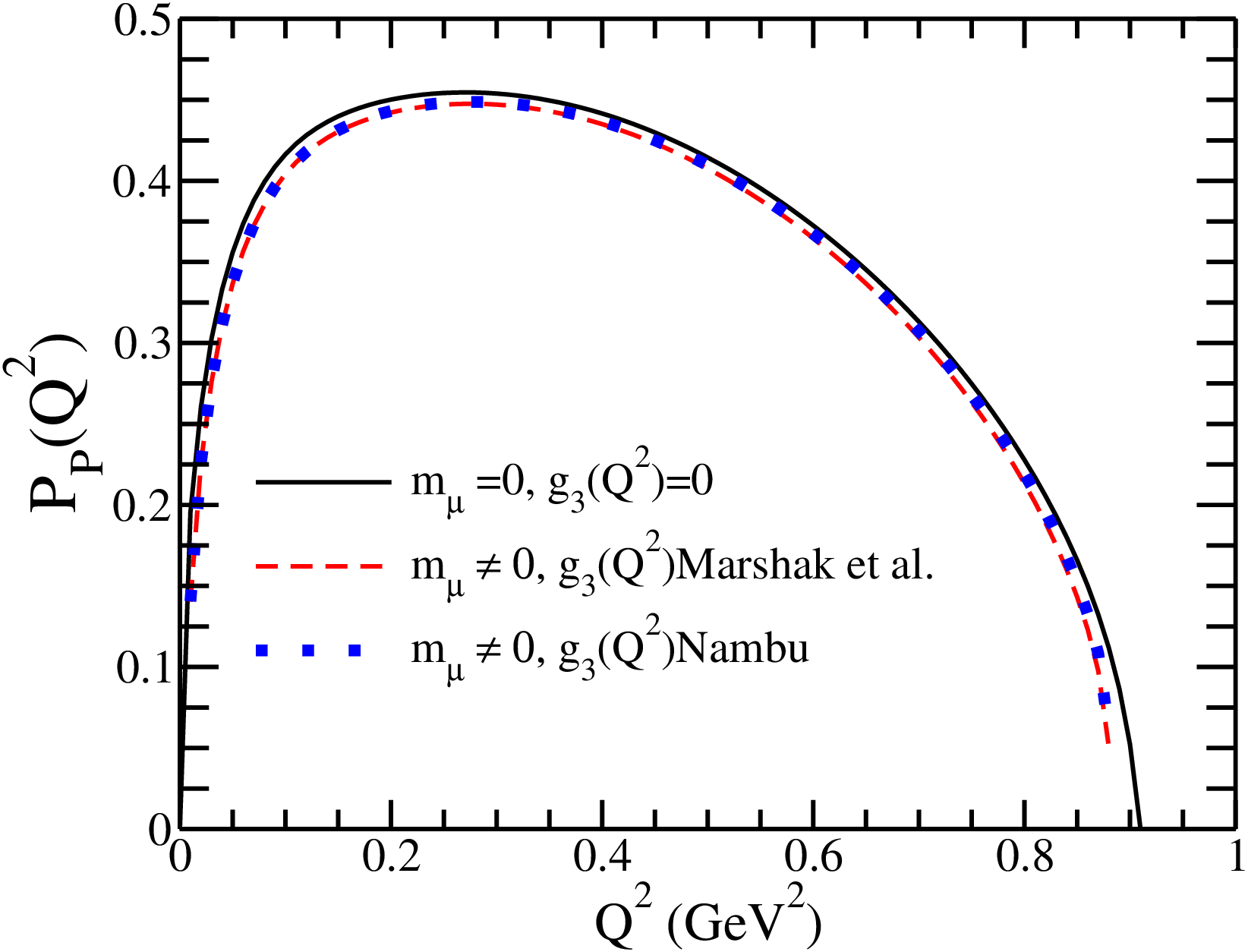}
   \caption{$\frac{d\sigma}{dQ^2},$ $ P_L(Q^2)$ and $P_P(Q^2)$ vs $Q^2$ at $E_{\bar \nu_{_\mu}}$= 1 GeV for $\bar\nu_\mu n \to \mu^+ \Sigma^-$ process.
   Lines and points have the same meaning as in Fig.~\ref{fp_vartn}.}\label{fp_vartn_sig}
\end{figure}

\begin{figure}
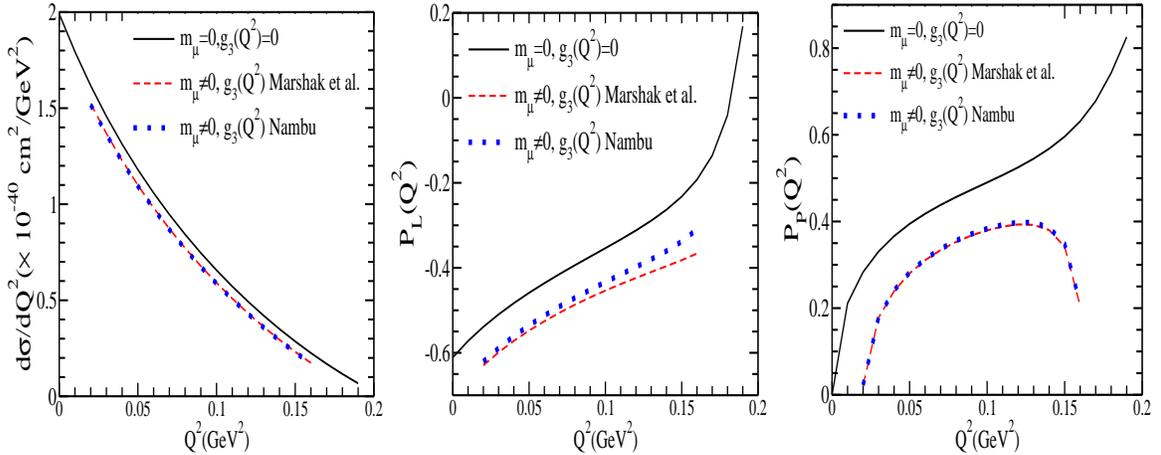

\includegraphics[height=6cm,width=5cm]{q2_sigma_minus_enu500MeV.eps}
\includegraphics[height=6cm,width=5cm]{sL_sigma_minus_enu500MeV.eps}
\includegraphics[height=6cm,width=5cm]{sT_sigma_minus_enu500MeV.eps}
\caption{$\frac{d\sigma}{dQ^2}$, $P_L(Q^2)$ and $P_P(Q^2)$ vs $Q^2$ for the process $\bar \nu_\mu n \to\mu^+
\Sigma^-$ at $E_{\bar \nu_{_\mu}}=0.5$ GeV. Lines and points have the same meaning as in Fig.~\ref{fp_vartn}. }
\label{sig_minus_500MeV}
\end{figure}

 We have made an attempt to explore the possibility of determining the pseudoscalar form factor $g_3^{NY}(Q^2)$ in 
 $|\Delta S|=1$ sector by including two models for $g_3^{NY}(Q^2)$ based on PCAC  
 and the corresponding Goldberger--Treiman relation in the strangeness sector using the parameterizations
 given in Eqs.~\ref{g3_marshak}(Marshak et al.~\cite{Marshak}) and \ref{g3_kaon_pole}(Nambu~\cite{Nambu:1960xd}). In Figs.~\ref{fp_vartn} and \ref{fp_vartn_sig},
 we show the effect of $g_3^{NY}(Q^2)$ on $\frac{d\sigma}{dQ^2}$,  $P_L(Q^2)$ and $P_P(Q^2)$ calculated for the processes
  $\bar\nu_\mu p \to \mu^+\Lambda$ and $\bar\nu_\mu n\to \mu^+ \Sigma^-$, respectively, at $E_{\bar \nu_{_\mu}}$=1 GeV. We see from 
  Figs.~\ref{fp_vartn} and \ref{fp_vartn_sig} that at $E_{\bar \nu_{_\mu}}=1$ GeV, sensitivity of the cross section $\frac{d\sigma}{dQ^2}$, or the
  polarization observables $P_L(Q^2)$ and $P_P(Q^2)$ to the pseudoscalar form factor, $g_3^{NY}(Q^2)$ is quite small. However, 
  at smaller antineutrino energies like $E_{\bar\nu_{_\mu}}$=0.5 GeV, the polarization components $P_L(Q^2)$ and $P_P(Q^2)$ 
  are quite sensitive to the value of the pseudoscalar form factor as shown in Figs.~\ref{lam_500MeV} and \ref{sig_minus_500MeV}.
It seems, therefore, possible in principle, to determine the pseudoscalar form factor in the hyperon polarization measurements at lower 
  energies relevant for the MicroBooNE~\cite{Chen:2007ae} and T2K~\cite{Abe:2015ibe} flux of antineutrinos. 

\begin{figure}
\includegraphics[height=6cm,width=5cm]{q2_lam_enu500MeV.eps}
\includegraphics[height=6cm,width=5cm]{sL_lam_enu500MeV.eps}
\includegraphics[height=6cm,width=5cm]{sT_lam_enu500MeV.eps}
\caption{$\frac{d\sigma}{dQ^2}$, $P_L(Q^2)$ and $P_P(Q^2)$ vs $Q^2$ for the process $\bar \nu_\mu p \to
\mu^+ \Lambda$ at $E_{\bar \nu_{_\mu}}=0.5$ GeV. Lines and points have the same meaning as in Fig.~\ref{fp_vartn}.}
\label{lam_500MeV}
\end{figure}
%

\subsection{Differential cross section $\frac{d\sigma}{dQ^2}$ and polarization components $P_L(Q^2)$ and $P_P(Q^2)$ for nuclear target}

\begin{figure}
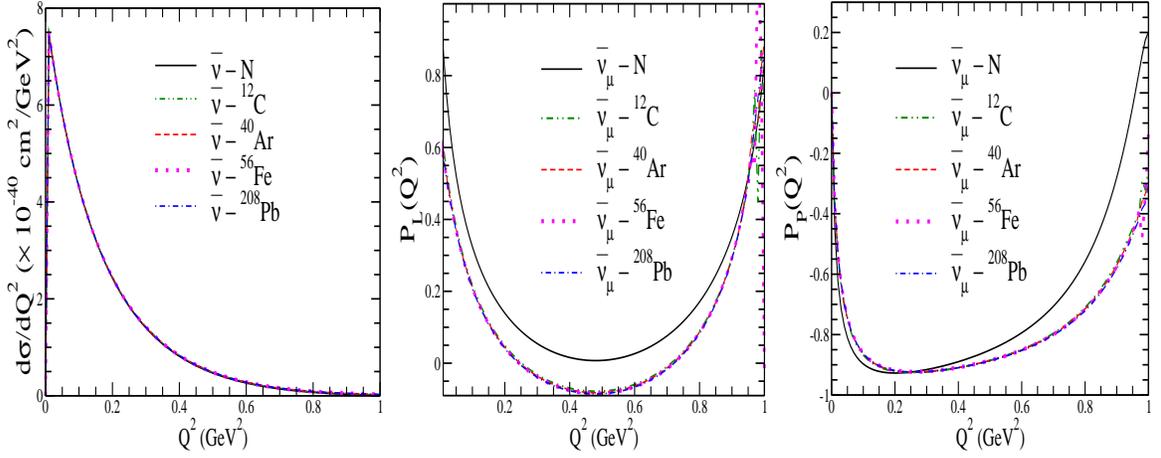

   \includegraphics[height=6cm,width=5cm]{q2_nucleus_f1f2fa_aml_lambda_enu1.eps}
   \includegraphics[height=6cm,width=5cm]{PL_nucleus_f1f2fa_aml_lambda_enu1.eps}
   \includegraphics[height=6cm,width=5cm]{Pp_nucleus_f1f2fa_aml_lambda_enu1.eps}
   \caption{ $\frac{d\sigma}{dQ^2}$, $P_L(Q^2)$ and $P_P(Q^2)$ vs $Q^2$ for the process $\bar \nu_\mu p \to \mu^+ \Lambda$ at $E_{\bar \nu_{_\mu}}$ = 1 GeV 
   for free nucleon(solid line) and different nuclei per interacting particle viz. $^{12}C$(dashed-double dotted), $^{40}Ar$(dashed line), $^{56}Fe$(dotted line) and $^{208}Pb$(dashed-dotted line)
   with  $m_\mu\ne0 $, $M_A$= 1.026 GeV.
   We have used $f_1^{p\Lambda}(Q^2),~f_2^{p\Lambda}(Q^2)~ \text{and} ~g_1^{p\Lambda}(Q^2)$ from Table~\ref{tab:formfac} and BBBA05 parameterization for nucleon form factors.}\label{Ma_dstbn_nucleus_lam}
  
\end{figure}

\begin{figure}
   \includegraphics[height=6cm,width=5cm]{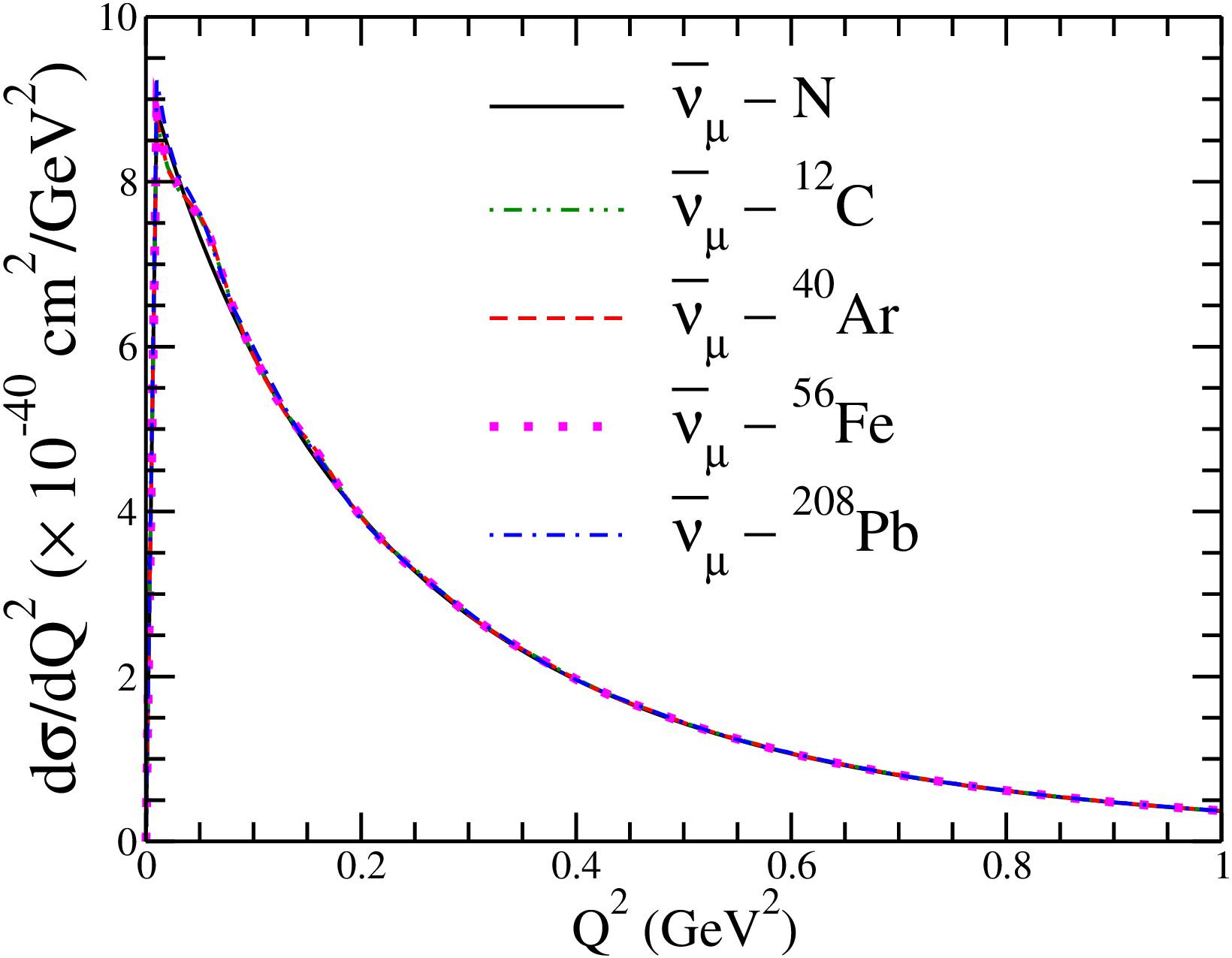}
   \includegraphics[height=6cm,width=5cm]{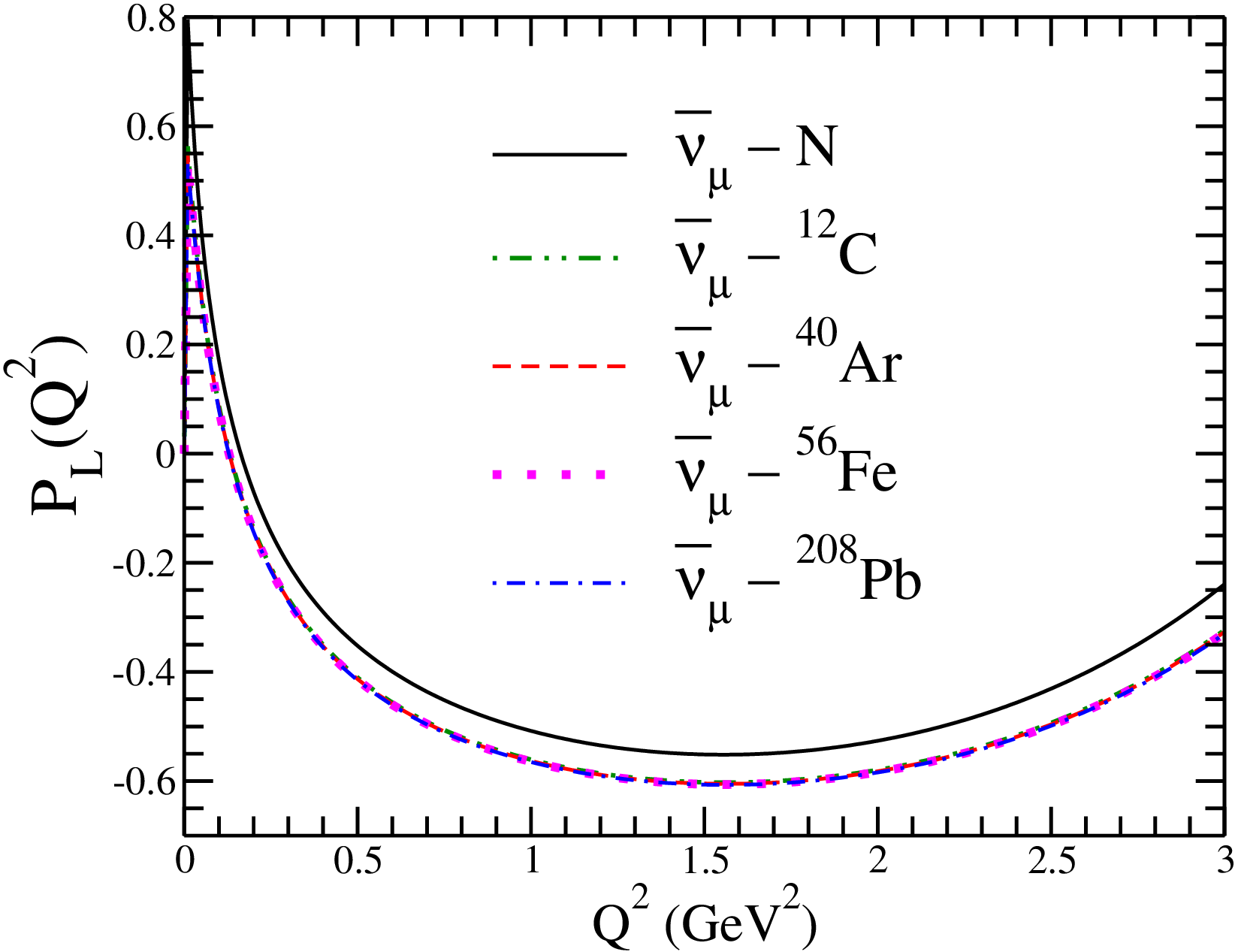}
   \includegraphics[height=6cm,width=5cm]{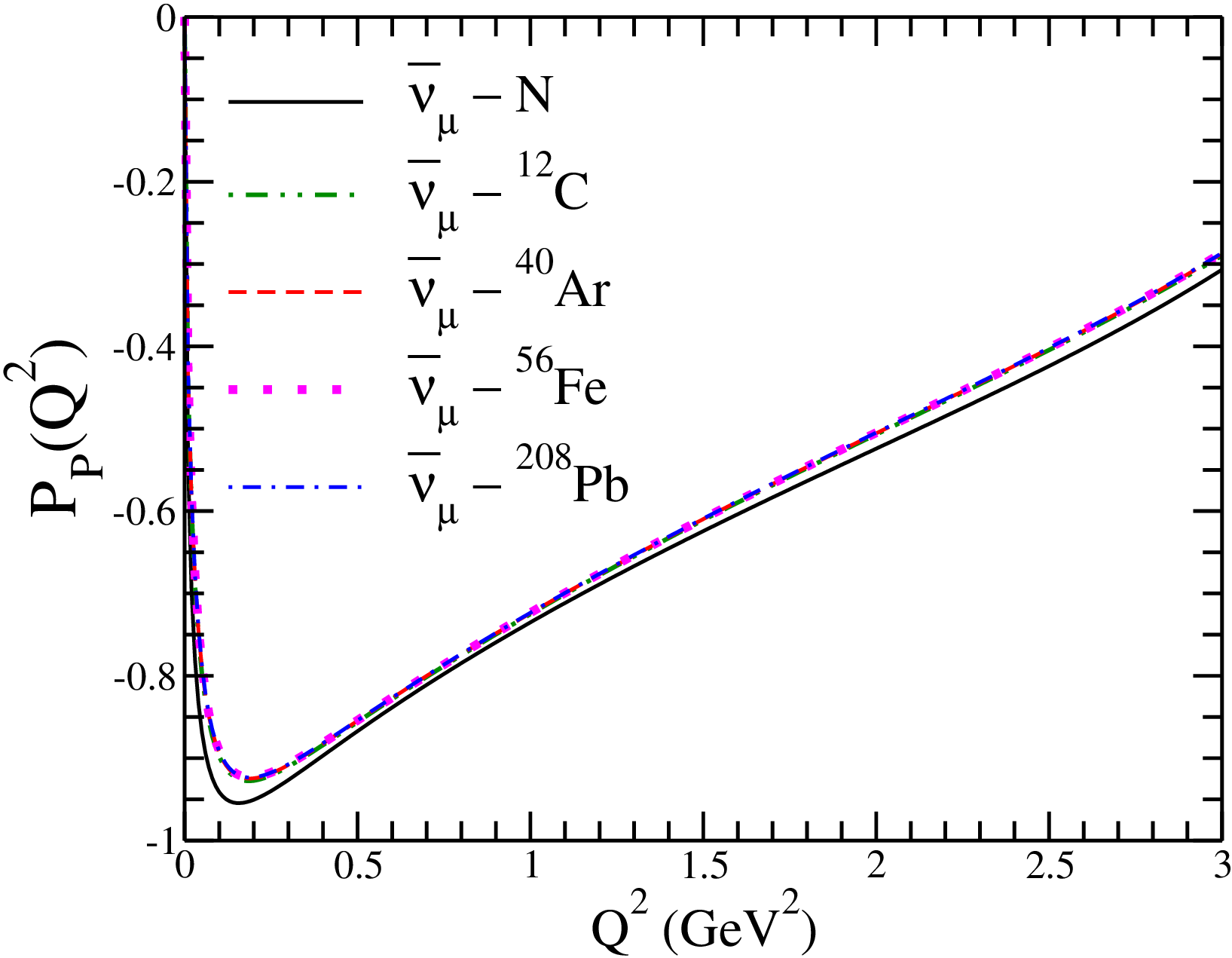}
   \caption{ $\frac{d\sigma}{dQ^2}$, $P_L(Q^2)$ and $P_P(Q^2)$ vs $Q^2$ for the process $\bar \nu_\mu p \to \mu^+ \Lambda$ at $E_{\bar \nu_{_\mu}}$ = 3 GeV.
    Lines and points have the same meaning as Fig.~\ref{Ma_dstbn_nucleus_lam}.}\label{Ma_dstbn_nucleus_lam_enu3}
\end{figure}

\begin{figure}
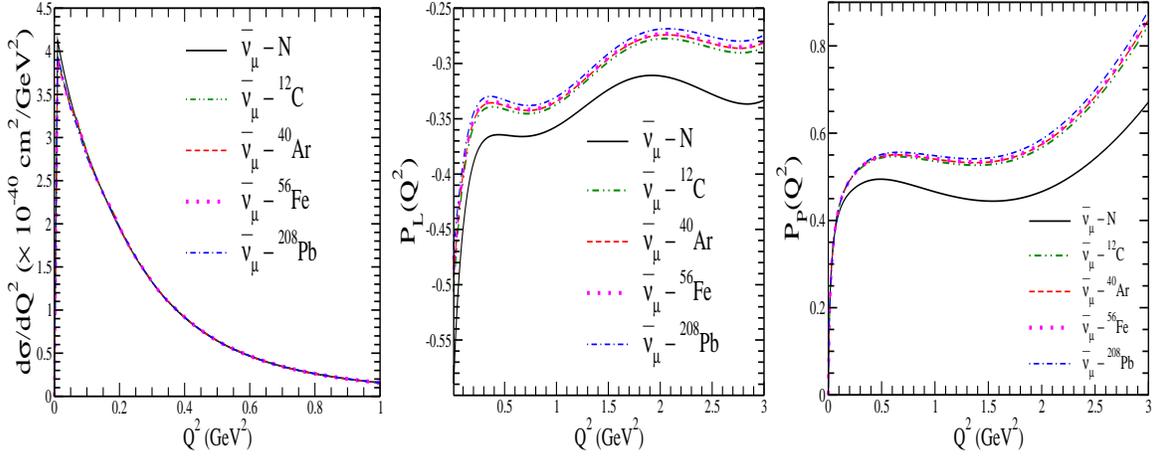

   \includegraphics[height=6cm,width=5cm]{q2_nucleus_f1f2fa_aml_sig_minus_enu3.eps}
   \includegraphics[height=6cm,width=5cm]{PL_nucleus_f1f2fa_aml_sig_minus_enu3.eps}
   \includegraphics[height=6cm,width=5cm]{Pp_nucleus_f1f2fa_aml_sig_minus_enu3.eps}
   \caption{ $\frac{d\sigma}{dQ^2}$, $P_L(Q^2)$ and $P_P(Q^2)$ vs $Q^2$ for the process $\bar \nu_\mu n \to \mu^+ \Sigma^-$ at $E_{\bar \nu_{_\mu}}$ = 3 GeV .
    Lines and points have the same meaning as Fig.~\ref{Ma_dstbn_nucleus_lam}. }\label{Ma_dstbn_nucleus_sig_enu3}
\end{figure}

In Figs.~\ref{Ma_dstbn_nucleus_lam}, \ref{Ma_dstbn_nucleus_lam_enu3}, \ref{Ma_dstbn_nucleus_sig} and \ref{Ma_dstbn_nucleus_sig_enu3}, we present the results in nuclei for differential cross section
$\frac{d\sigma}{dQ^2}$, longitudinal ($P_L(Q^2)$) and perpendicular ($P_P(Q^2)$) components of $\Lambda$ and $\Sigma$ polarization at $E_{\bar{\nu}_\mu}=$ 1 and 3 GeV
 for various nuclei like $^{12}$C, $^{40}$Ar, $^{56}$Fe, and $^{208}$Pb using
 Eqs.~\ref{diffnuc}, \ref{PL_nuc} and \ref{Pp_nuc}. The results are compared with the results for the free nucleon case. We find that at $E_{\bar {\nu}_{\mu}}$ = 1 GeV, the differential 
 cross section $\frac{d\sigma}{dQ^2}$ hardly changes with the inclusion of nuclear medium effects. This is in contrast to the quasielastic 
 reaction $\nu_l(\bar \nu_l)~+~n(p) ~\to~ l^-(l^+)~+~p(n)$. This 
 is due to the lack of any Pauli blocking of the momentum of the final hyperon which has its own Fermi sea.
 The polarization observables $P_L(Q^2)$ and $P_P(Q^2)$  show some dependence on nuclear medium effects. The nature of this dependence 
 is different for $P_L(Q^2)$ and $P_P(Q^2)$ as
 well as it is different for $\Lambda$ and $\Sigma$ hyperons. For example, in the case of $\bar\nu_\mu p \to \mu^+ \Lambda$, the result for 
 $P_L(Q^2)$ at low $Q^2$ is hardly affected by nuclear medium effects, however, with the increase in $Q^2$ the effect of nuclear medium increases.
 The effect becomes maximum for $Q^2~\sim$ 0.5 GeV$^2$ and then decreases with further 
 increase in $Q^2$. While in the case of $P_P(Q^2)$ the effect is smaller as compared to $P_L(Q^2)$ i.e. almost negligible for $Q^2 ~<$  0.4 GeV$^2$ and a slight increase for $Q^2~>$  0.4 GeV$^2$.

For $\bar\nu_\mu n \to \mu^+ \Sigma^-$, the difference in the results obtained for nucleon and nuclear targets
 increases with the increase in $Q^2$, 
 both for $P_L(Q^2)$ and $P_P(Q^2)$. Furthermore, we find that there is very little nuclear mass number(A) dependence of nuclear medium effects. Moreover, the nuclear effect becomes smaller 
 with the increase in antineutrino energy.

\begin{figure}
   \includegraphics[height=6cm,width=5cm]{q2_nucleus_f1f2fa_aml_sig_minus_enu1.eps}
   \includegraphics[height=6cm,width=5cm]{PL_nucleus_f1f2fa_aml_sig_minus_enu1.eps}
   \includegraphics[height=6cm,width=5cm]{Pp_nucleus_f1f2fa_aml_sig_minus_enu1.eps}
   \caption{ $\frac{d\sigma}{dQ^2}$, $P_L(Q^2)$ and $P_P(Q^2)$ vs $Q^2$ for the process $\bar \nu_\mu n \to \mu^+ \Sigma^-$ at $E_{\bar \nu_{_\mu}}$ = 1 GeV.
   Lines and points have the same meaning as Fig.~\ref{Ma_dstbn_nucleus_lam}.}\label{Ma_dstbn_nucleus_sig}
\end{figure}

\subsection{Flux averaged differential cross section and polarization components}  

\begin{figure}
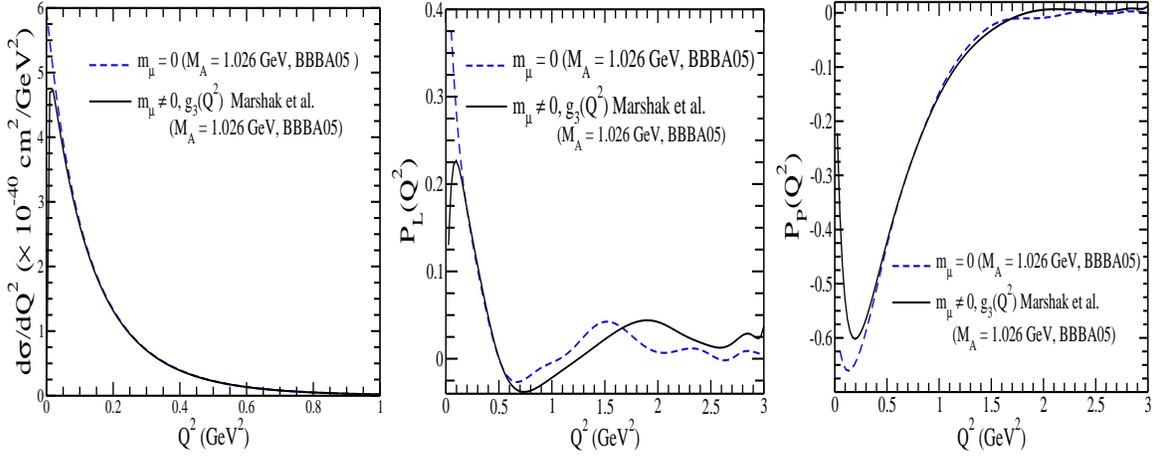

   \includegraphics[height=6cm,width=5cm]{q2_hyp_MicroBooNE_Ar.eps}
   \includegraphics[height=6cm,width=5cm]{sL_lam_MicroBooNE_Ar.eps}
   \includegraphics[height=6cm,width=5cm]{sT_lam_MicroBooNE_Ar.eps}
   \caption{ $\frac{d\sigma}{dQ^2}$, $P_L(Q^2)$ and $P_P(Q^2)$ vs $Q^2$ for the process
   $\bar \nu_\mu p \to \mu^+ \Lambda$($^{40}$Ar target) averaged over the MicroBooNE~\cite{Chen:2007ae} spectrum, 
   using $f_1^{p\Lambda}(Q^2),~f_2^{p\Lambda}(Q^2),~g_1^{p\Lambda}(Q^2)$ 
   from Table-\ref{tab:formfac} and the BBBA05 parameterization~\cite{Bradford:2006yz} for the nucleon form factors with $m_\mu=0$ and $M_A=1.026$ GeV(dashed line),
    and $m_\mu \ne0$, $M_A=1.026$ GeV with $g_3^{p\Lambda}(Q^2)$ from 
   Marshak et al.~\cite{Marshak}(solid line).}\label{MicroBooNE}
\end{figure}

\begin{figure}
   \includegraphics[height=6cm,width=5cm]{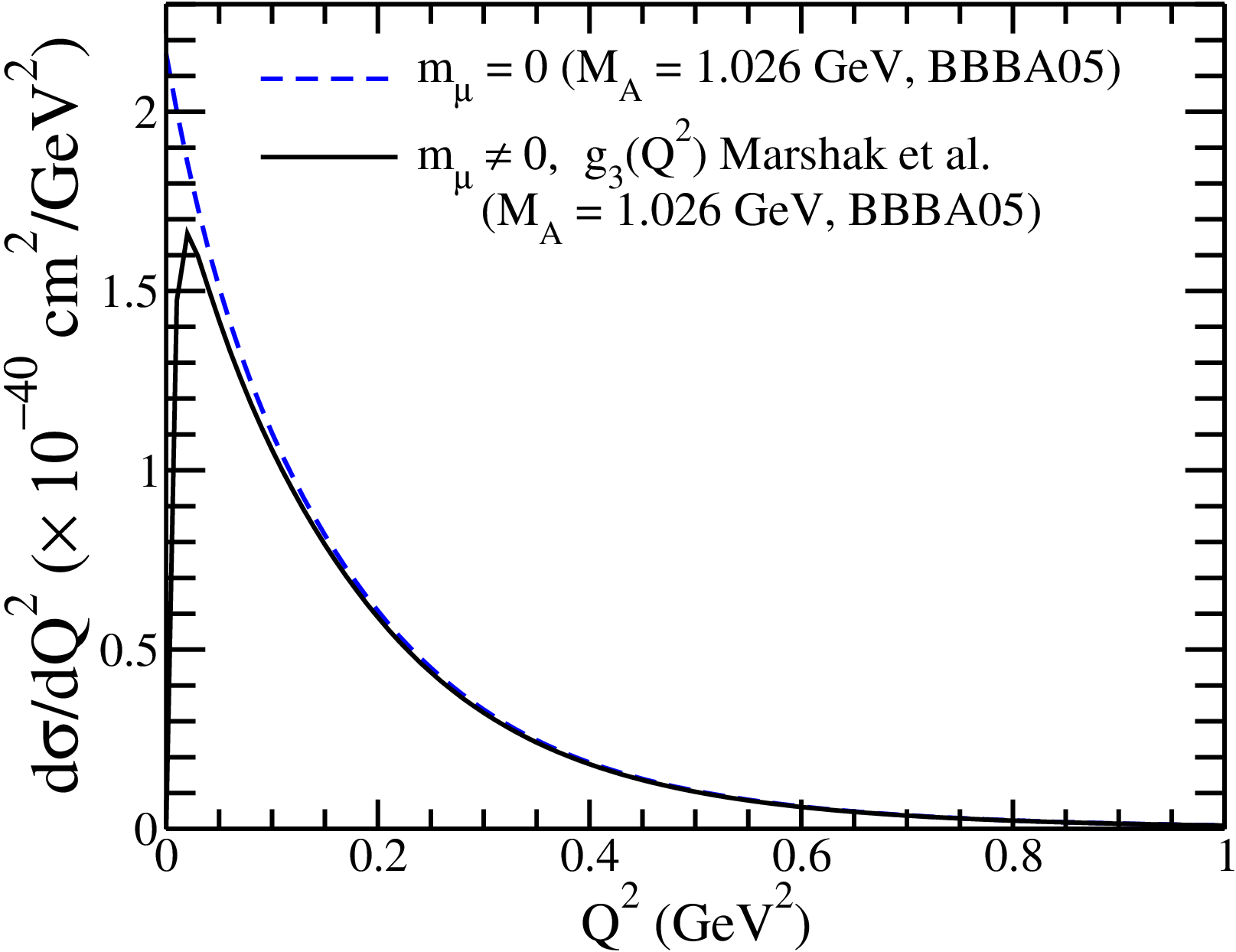}
   \includegraphics[height=6cm,width=5cm]{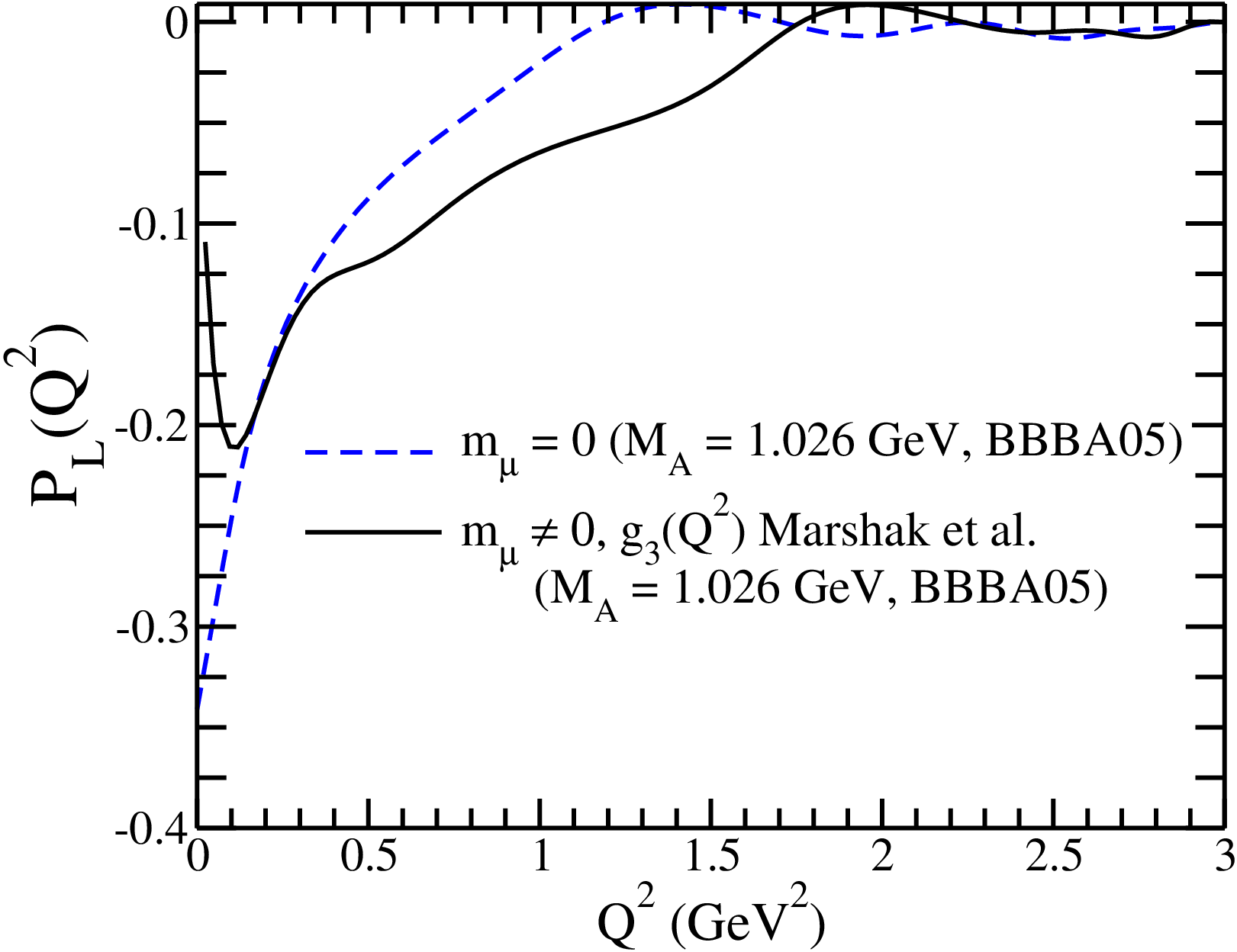}
   \includegraphics[height=6cm,width=5cm]{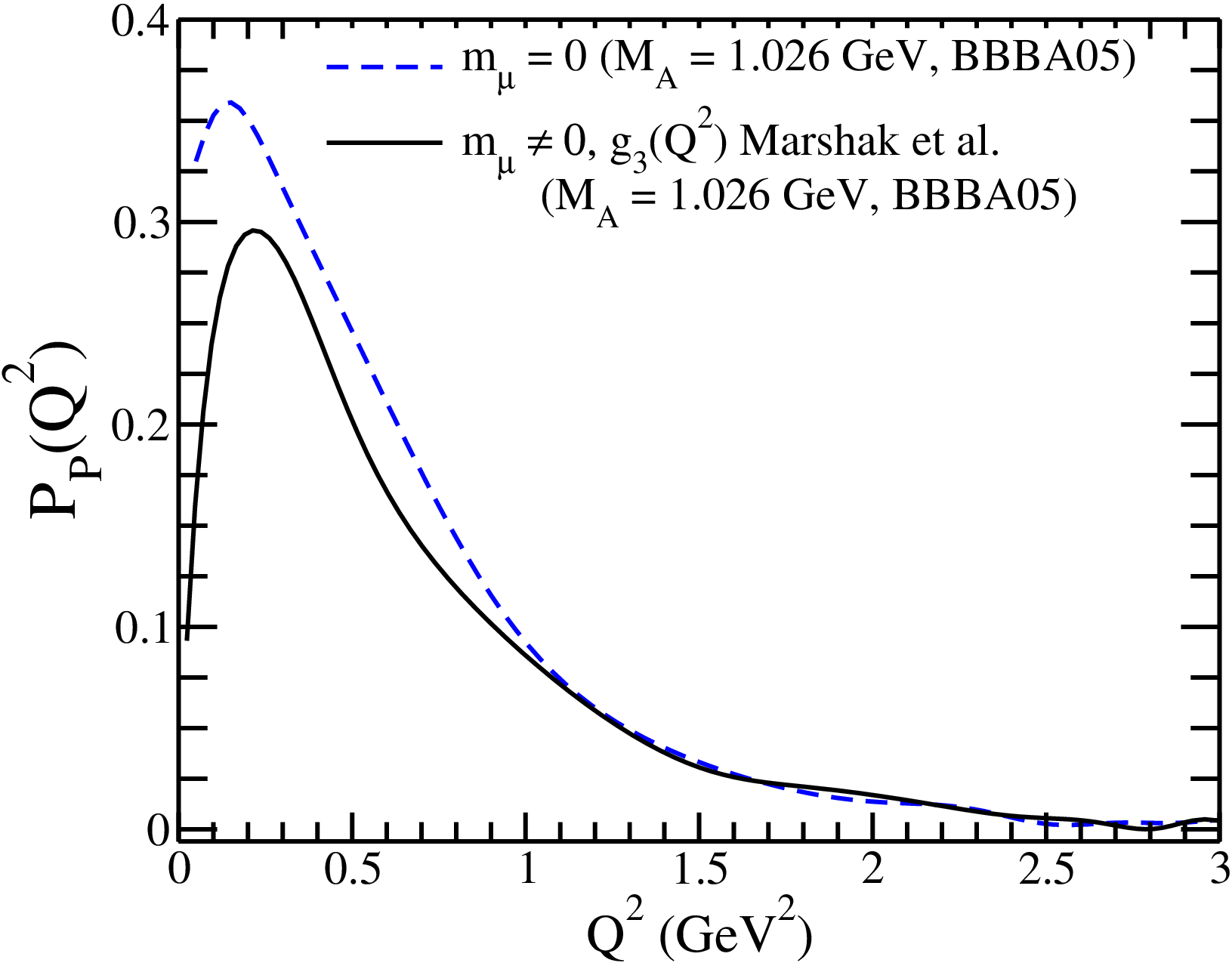}
   \caption{ $\frac{d\sigma}{dQ^2}$, $P_L(Q^2)$ and $P_P(Q^2)$ vs $Q^2$ for the process
   $\bar \nu_\mu n \to \mu^+ \Sigma^-$($^{40}$Ar target) averaged over MicroBooNE~\cite{Chen:2007ae} spectrum. Lines and points have 
   the same meaning as in Fig.~\ref{MicroBooNE}. }\label{sig_minus_mcrbne}
\end{figure}

  Currently, there are some neutrino experiments which are making measurements on neutrino--nucleus cross 
  sections~\cite{Fields:2013zhk,Abe:2015ibe,Palamara:2011zz}. The LArTPC detector proposed for 
MicroBooNE~\cite{Chen:2007ae}, ArgoNeut~\cite{Palamara:2011zz}, LAr1-ND, ICARUS-T600~\cite{combined} and DUNE~\cite{Acciarri:2015uup} may be able to measure
the tracks corresponding to nucleon and pion coming from $\Lambda$ decay. A measurement of the 
  asymmetry in the angular distribution of pions will give information about the hyperon ($\Lambda,\Sigma^-$) 
  polarization. For the purpose of analyzing these experiments, we have convoluted $\frac{d\sigma}{dQ^2}$ and $P_{L,P}(Q^2)$
distributions over the flux $\Phi(E_{\bar \nu_{_\mu}})$ available for different experiments using the expression given by,
\begin{equation}
\langle F(Q^2) \rangle = \frac{\int_ {E_{th}}^{E_{max}} F(Q^2,E_{\bar \nu_{_\mu}}) \Phi(E_{\bar \nu_{_\mu}}) d E_{\bar \nu_{_\mu}}   }{\int_{E_{min}}^{E_{max}}
\Phi(E_{\bar \nu_{_\mu}}) d E_{\bar \nu_{_\mu}}},
\end{equation}
where the function $F(Q^2,E_{\bar \nu_{_\mu}})$ represents $\frac{d\sigma}{dQ^2}(Q^2,E_{\bar \nu_{_\mu}})$, $P_{L}(Q^2,E_{\bar \nu_{_\mu}})$ and 
$P_{P}(Q^2,E_{\bar \nu_{_\mu}})$ given in Eqs.~\ref{diffnuc},~\ref{PL_nuc} and \ref{Pp_nuc} respectively. $E_{th}$, $E_{min}$, $E_{max}$ are the threshold energy 
and the minimum and maximum energies of the 
antineutrino fluxes corresponding to these experiments. In Figs.~\ref{MicroBooNE} and \ref{sig_minus_mcrbne}, we have shown the flux averaged 
$\langle\frac{d\sigma}{dQ^2}\rangle$, $\langle P_L(Q^2)\rangle$ and $\langle P_P(Q^2)\rangle$ for reactions $\bar\nu_\mu p \to \mu^+\Lambda$ and $ \bar\nu_\mu n\to \mu^+ \Sigma^-$, respectively,
corresponding to the MicroBooNE~\cite{Chen:2007ae} antineutrino experiment in $^{40}$Ar using $M_A$=1.026 GeV and $g^{NY}_3(Q^2)\ne0$.
  
\begin{figure}
   \includegraphics[height=6cm,width=5cm]{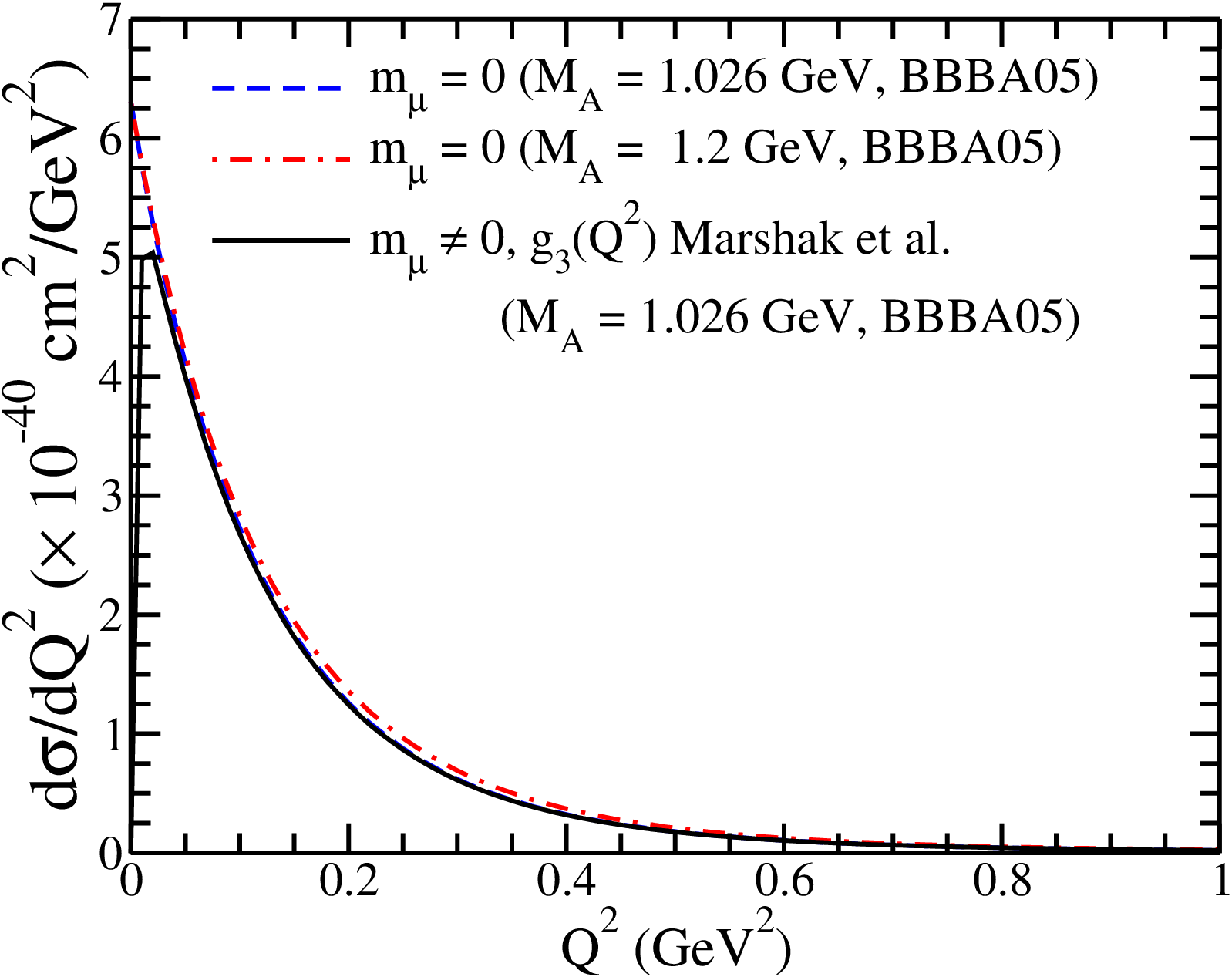}
   \includegraphics[height=6cm,width=5cm]{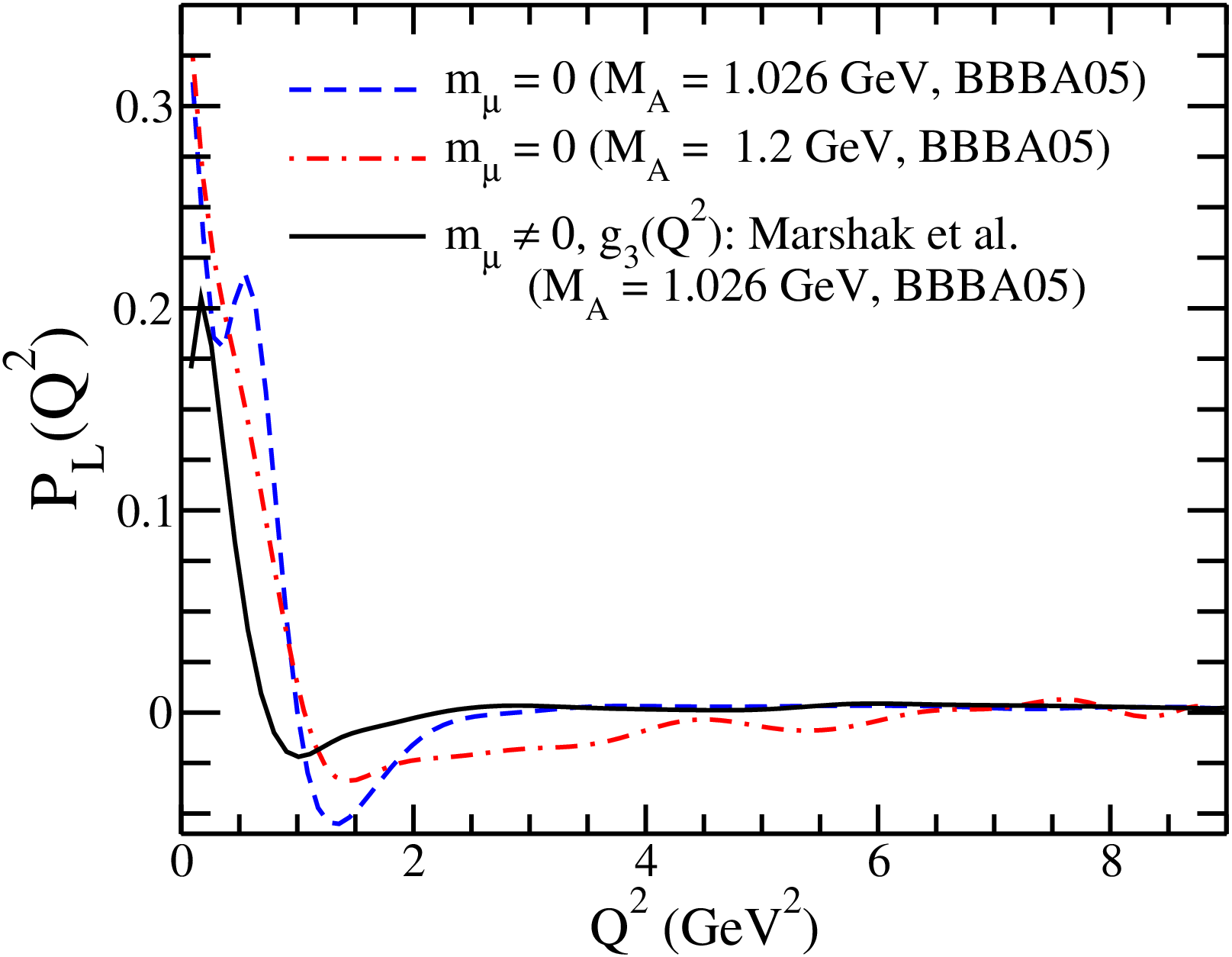}
   \includegraphics[height=6cm,width=5cm]{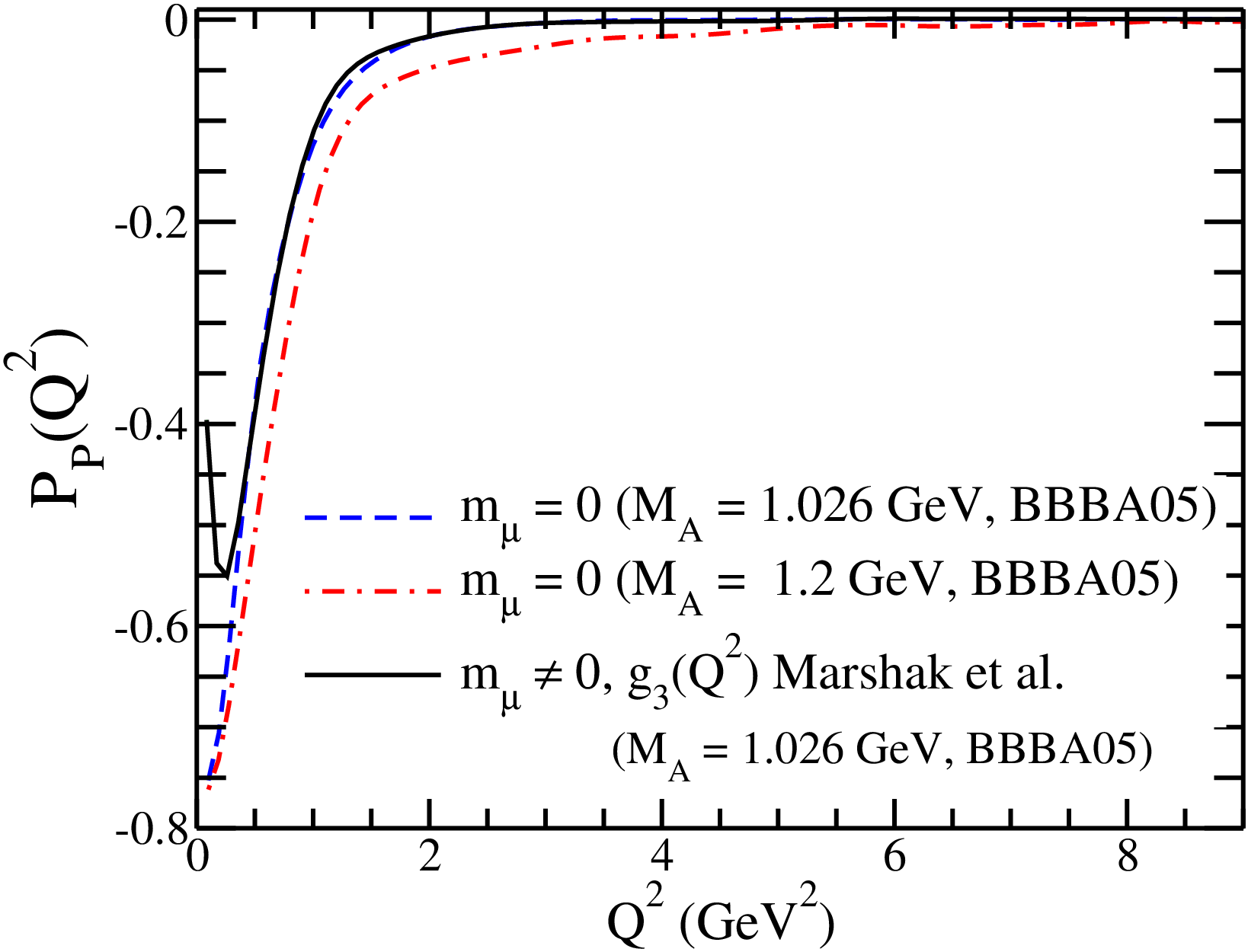}
   \caption{ $\frac{d\sigma}{dQ^2}$, $P_L(Q^2)$ and $P_P(Q^2)$ vs $Q^2$ for the process $\bar \nu_\mu p \to \mu^+ \Lambda$($^{12}$C target) averaged over T2K~\cite{Abe:2015ibe} spectrum,
   using $f_1^{p\Lambda}(Q^2),~f_2^{p\Lambda}(Q^2),~g_1^{p\Lambda}(Q^2)$ 
   from Table-\ref{tab:formfac} and the BBBA05 parameterization~\cite{Bradford:2006yz} for the nucleon form factors with $m_\mu=0$ and $M_A=1.026$ GeV(dashed line), $m_\mu=0$ and 
   $M_A=1.2$ GeV(dashed--dotted line) and $m_\mu \ne0$, $M_A=1.026$ GeV with $g_3^{p\Lambda}(Q^2)$ from 
   Marshak et al.~\cite{Marshak}(solid line).}\label{lam_T2K}
\end{figure}

\begin{figure}
   \includegraphics[height=6cm,width=5cm]{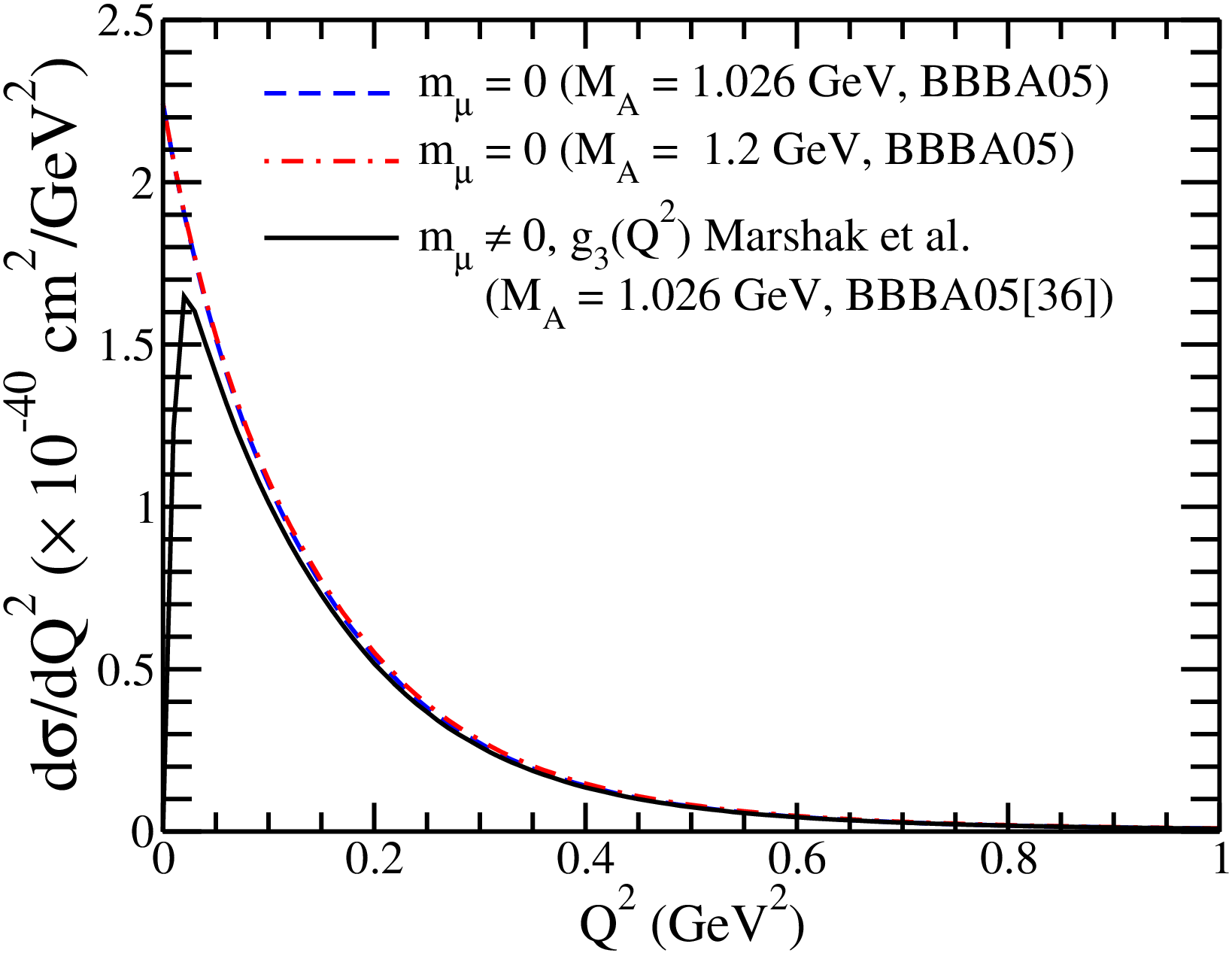}
   \includegraphics[height=6cm,width=5cm]{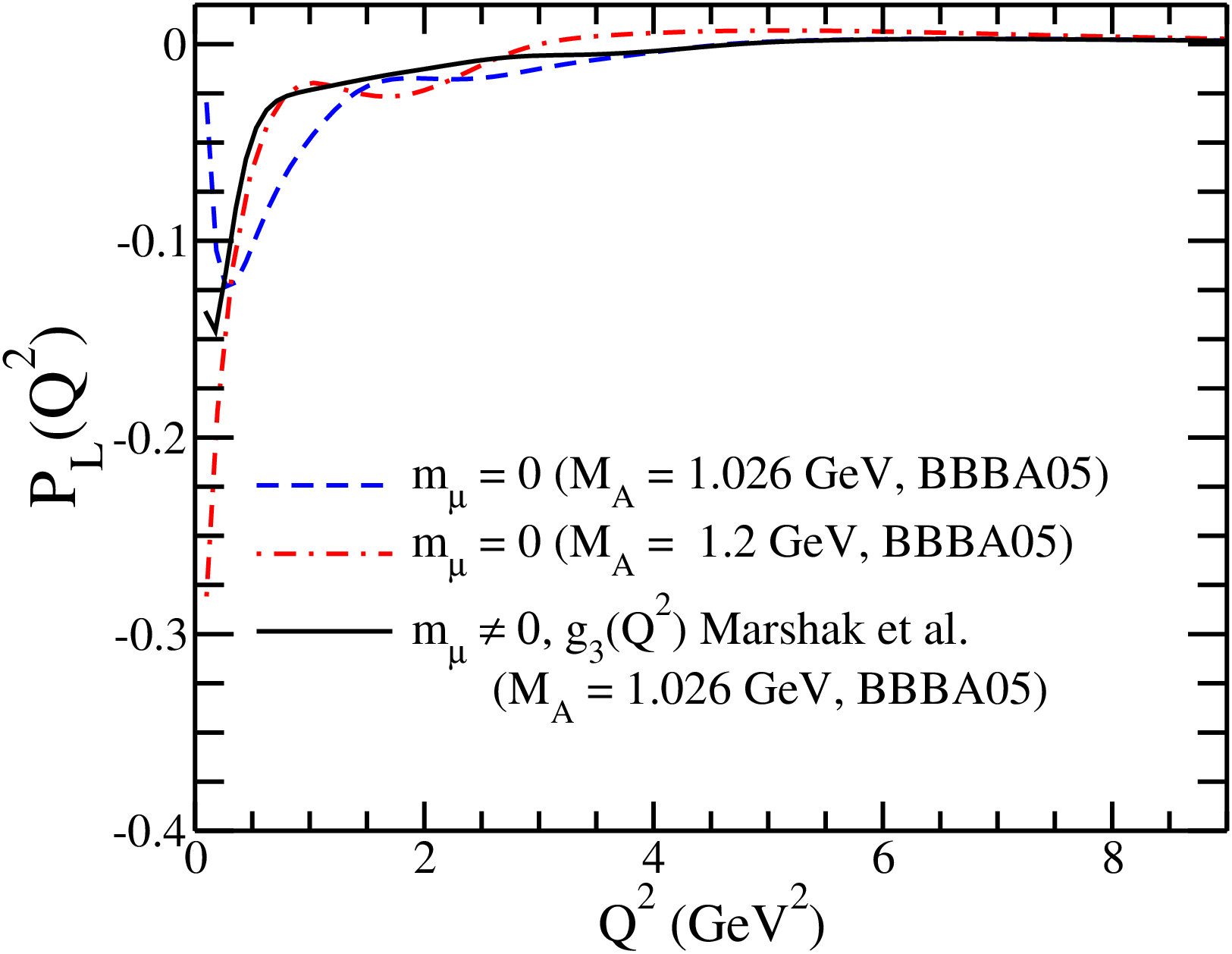}
   \includegraphics[height=6cm,width=5cm]{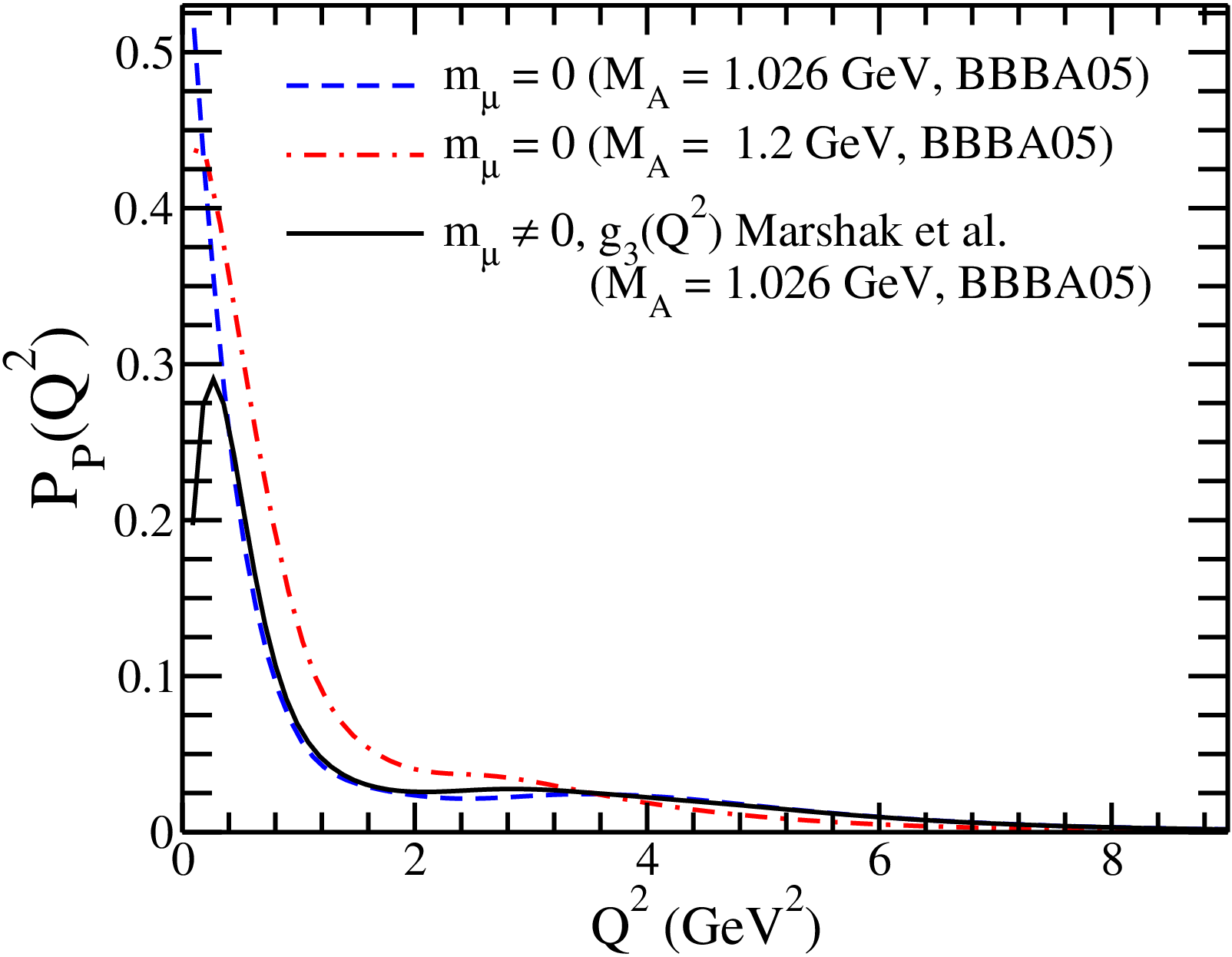}
   \caption{ $\frac{d\sigma}{dQ^2}$, $P_L(Q^2)$ and $P_P(Q^2)$ vs $Q^2$ for the process $\bar \nu_\mu n \to \mu^+ \Sigma^-$($^{12}$C target) averaged over T2K~\cite{Abe:2015ibe} spectrum.
   Lines and points have the same meaning as in Fig.~\ref{lam_T2K}.}\label{sig_minus_T2K}
\end{figure}

\begin{figure}
   \includegraphics[height=6cm,width=5cm]{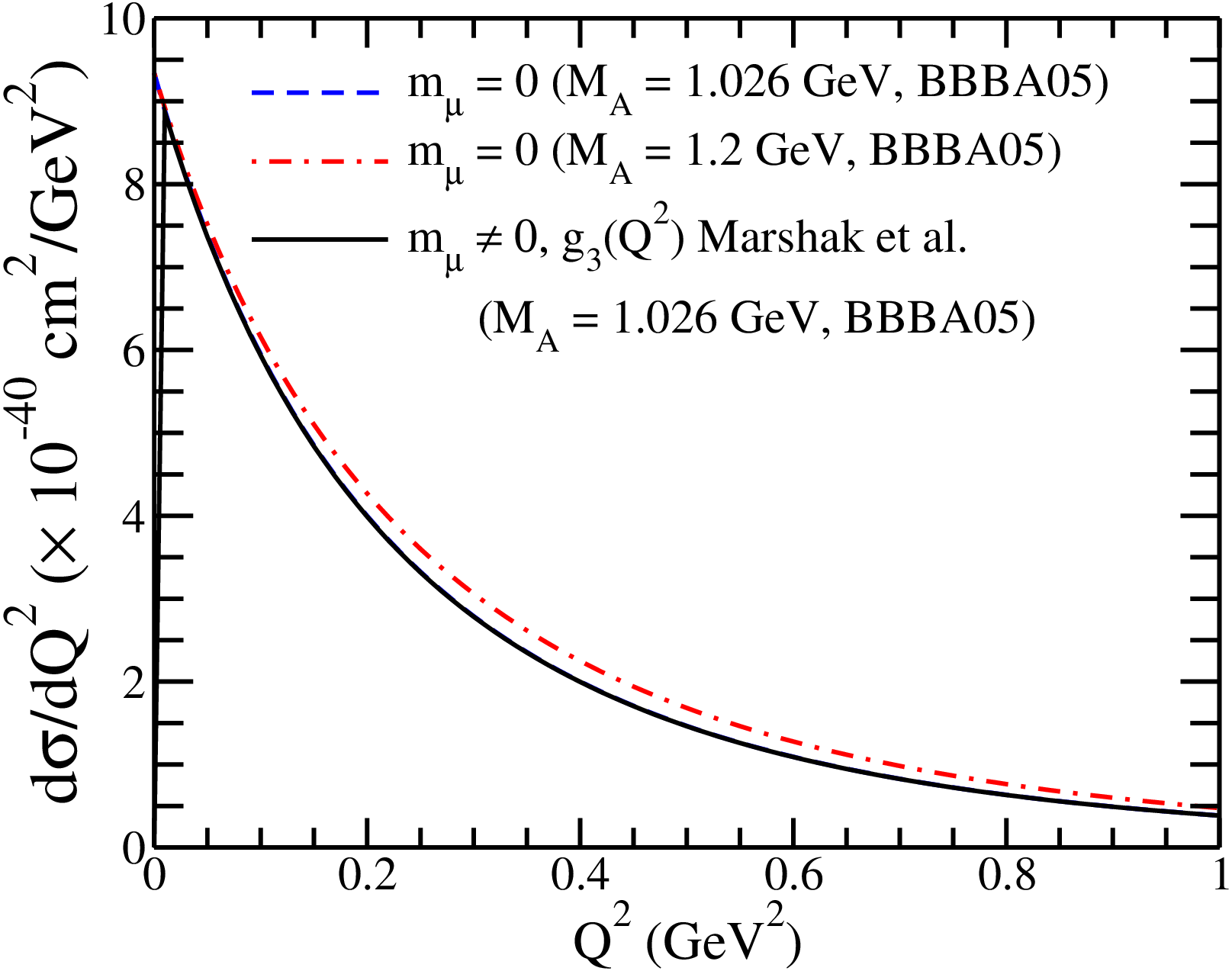}
   \includegraphics[height=6cm,width=5cm]{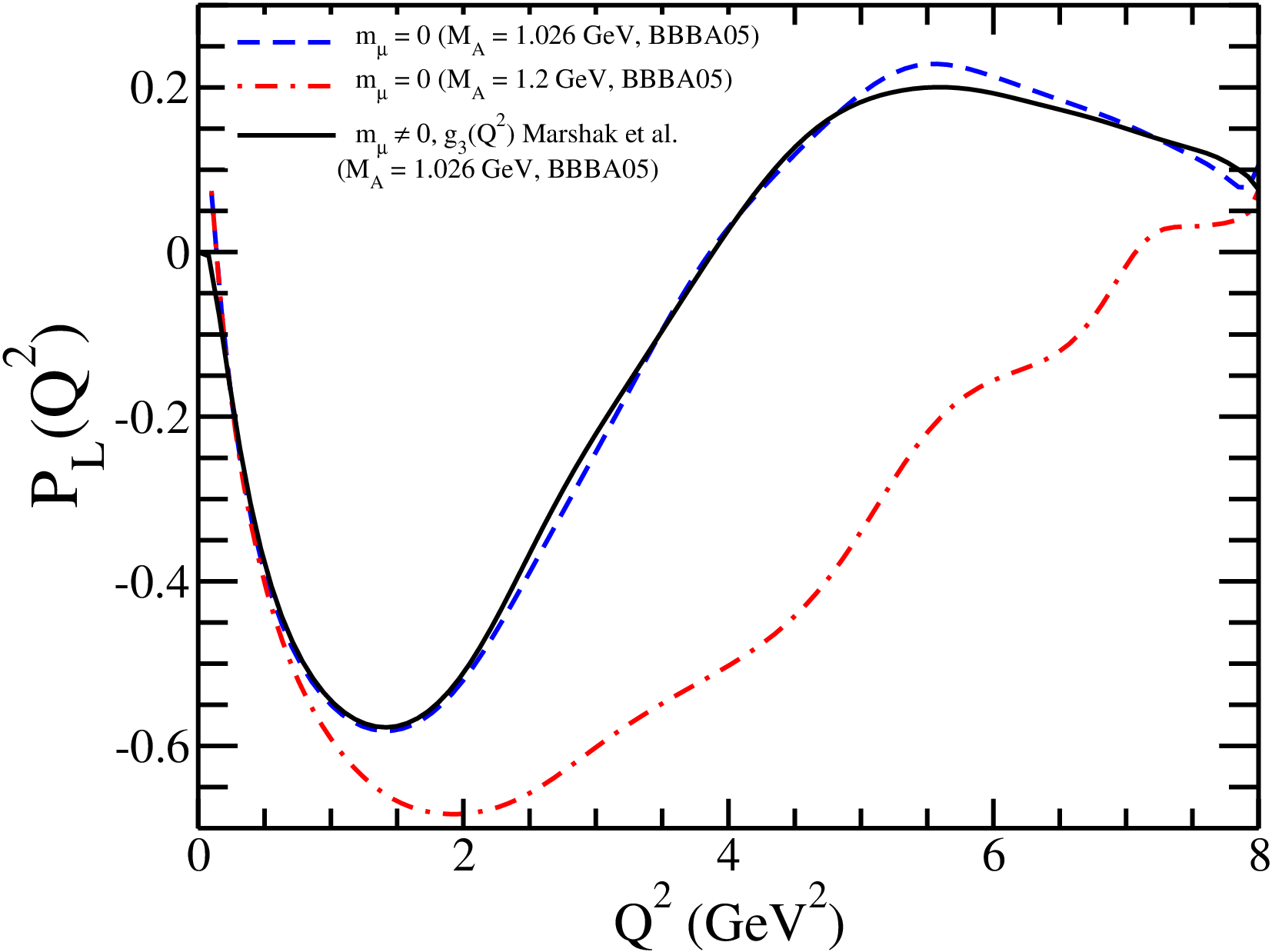}
   \includegraphics[height=6cm,width=5cm]{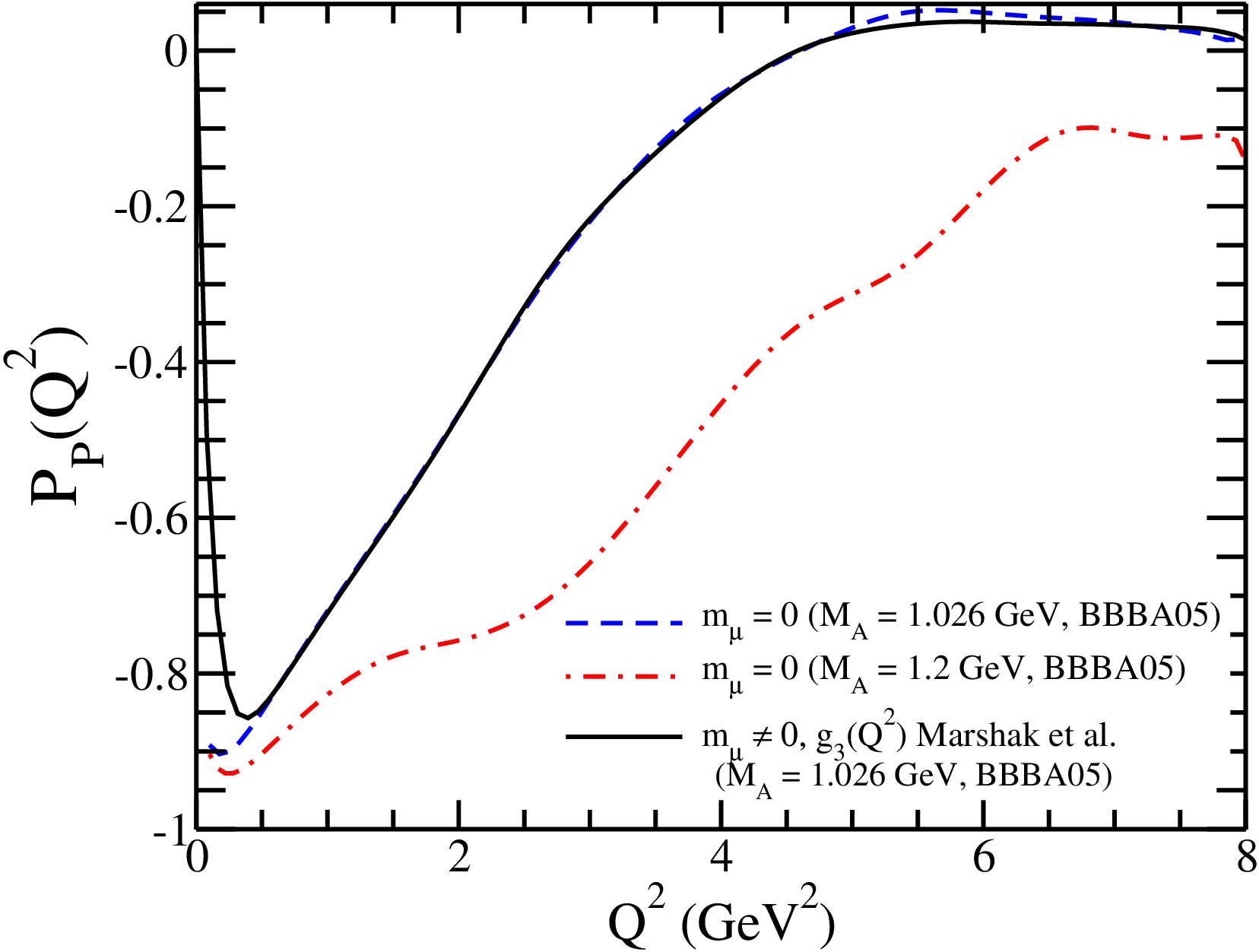}
   \caption{ $\frac{d\sigma}{dQ^2}$, $P_L(Q^2)$ and $P_P(Q^2)$ vs $Q^2$ for the process $\bar \nu_\mu p \to \mu^+ \Lambda$($^{208}$Pb target) averaged over MINER$\nu$A~\cite{Fields:2013zhk} spectrum. 
   Lines and points have 
   the same meaning as in Fig.~\ref{lam_T2K}. }\label{Minerva}
\end{figure}

   We have also shown in Figs.~\ref{lam_T2K} and \ref{sig_minus_T2K}, the flux averaged results of $\langle\frac{d\sigma}{dQ^2}\rangle$, $\langle P_L(Q^2)\rangle$ and $\langle P_P(Q^2)\rangle$ 
   for reactions $\bar\nu_\mu p \to \mu^+\Lambda$
   and $ \bar\nu_\mu n\to \mu^+ \Sigma^-$ respectively, for $^{12}$C target corresponding to the T2K~\cite{Abe:2015ibe} antineutrino spectrum. Similar results are presented
   for these reactions corresponding to
   MINER$\nu$A~\cite{Fields:2013zhk} experiment in $^{208}$Pb target for the
   antineutrino beam with average energy of 3.6 GeV in Figs.~\ref{Minerva} and \ref{sig_minus_minerva}.  
   It may be observed from these figures that polarization measurements on $\bar\nu_\mu p \to \mu^+\Lambda$ and $\bar \nu_\mu n \to \mu^+ \Sigma^-$ 
    in all these experiments will enable us to independently determine the value of axial vector form factor in the strangeness sector.
    
Moreover, at lower $\bar \nu_\mu$ energies relevant to MicroBooNE~\cite{Chen:2007ae} and T2K~\cite{Abe:2015ibe} experiments, it is also possible to determine the pseudoscalar from factors
and test the hypothesis of PCAC in the strangeness sector.

\begin{figure}
   \includegraphics[height=6cm,width=5cm]{q2_sig_minus_Minerva_Fe.eps}
   \includegraphics[height=6cm,width=5cm]{sL_sig_minus_Minerva_Fe.eps}
   \includegraphics[height=6cm,width=5cm]{sT_sig_minus_Minerva_Fe.eps}
   \caption{ $\frac{d\sigma}{dQ^2}$, $P_L(Q^2)$ and $P_P(Q^2)$ vs $Q^2$ for the process $\bar \nu_\mu n \to \mu^+ \Sigma^-$($^{208}$Pb target) averaged over MINER$\nu$A~\cite{Fields:2013zhk} spectrum. 
   Lines and points have the same meaning as in Fig.~\ref{lam_T2K}.}\label{sig_minus_minerva}
\end{figure}
\begin{figure}
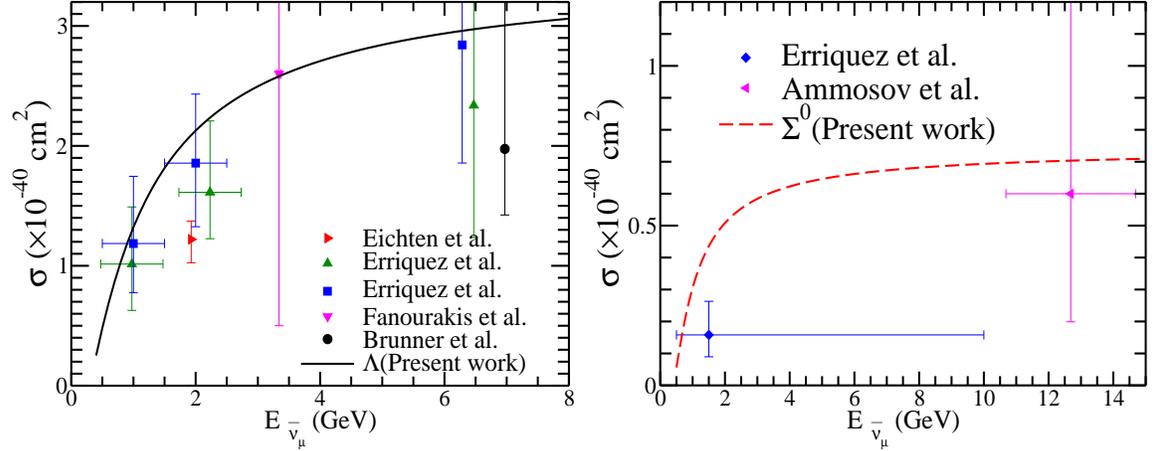

   \includegraphics[height=6cm,width=7.5cm]{sigma_vs_enu.eps}
   \includegraphics[height=6cm,width=7.5cm]{xsctn_sig0_vs_enu.eps}
   \caption{Theoretical curves for total cross section($\sigma$) vs $E_{\bar \nu_{_\mu}}$ corresponding to the processes $\bar \nu_\mu p \to \mu^+ \Lambda$(solid line) in the left panel
   and $\bar \nu_\mu p \to \mu^+ \Sigma^0$(dashed line) in the right panel
    using $f_1^{NY}(Q^2),~f_2^{NY}(Q^2),~g_1^{NY}(Q^2)$ 
   from Table-\ref{tab:formfac}, $g_3^{NY}(Q^2)$ from Marshak et al.~\cite{Marshak} given in Eq.~\ref{g3_marshak} with $M_A$ = 1.026 GeV. Experimental results for the process
   $\bar \nu_\mu p \to \mu^+ \Lambda$ (triangle right~\cite{Eichten:1972bb}, triangle up~\cite{Erriquez:1977tr}, square~\cite{Erriquez:1978pg},
 triangle down($\sigma = 2.6^{+5.9}_{-2.1} \times 10^{-40} cm^2$)~\cite{Fanourakis:1980si}, circle~\cite{Brunner:1989kw}) 
 and for the process $\bar \nu_\mu p \to \mu^+ \Sigma^0$ (diamond~\cite{Erriquez:1977tr})
are shown with error bars. } \label{xsec_enu}
\end{figure}
\begin{figure}
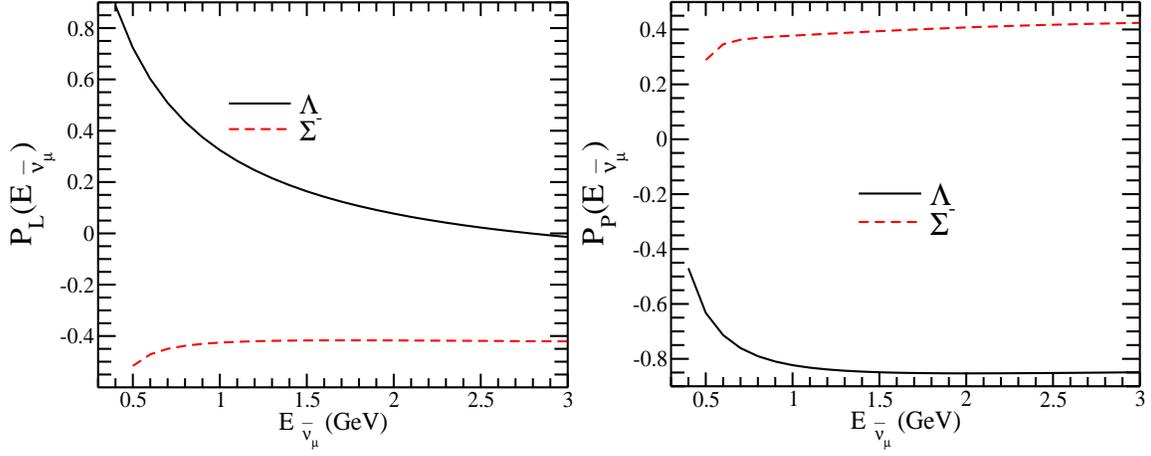

   \includegraphics[height=6cm,width=7.5cm]{sL_vs_enu.eps}
   \includegraphics[height=6cm,width=7.5cm]{sT_vs_enu.eps}
   \caption{Polarization components $P_L(E_{\bar \nu_{_\mu}})$ and $P_P(E_{\bar \nu_{_\mu}})$ vs $E_{\bar \nu_{_\mu}}$ using Eq.~\ref{energy_pol} for the processes 
   $\bar \nu_\mu p \to \mu^+ \Lambda$(solid line) 
   and $\bar \nu_\mu n \to \mu^+ \Sigma^-$(dashed line) using $f_1^{NY}(Q^2),~f_2^{NY}(Q^2),~g_1^{NY}(Q^2)$ 
   from Table-\ref{tab:formfac} and $g_3^{NY}(Q^2)$ from Marshak et al.~\cite{Marshak} with $M_A$ = 1.026 GeV.} \label{pol_enu}
\end{figure}
\subsection{Energy dependence of total cross section and average polarizations}
 We have calculated the total cross section $\sigma(E_{\bar \nu_{_\mu}})$ as a function of energy, given as:
 \begin{equation}
  \sigma(E_{\bar \nu_{_\mu}})=\int^{Q^2_{max}}_{Q^2_{min}} \frac{d\sigma}{dQ^2}(Q^2,E_{\bar \nu_{_\mu}}) dQ^2
 \end{equation}
 for $\bar \nu_\mu p
 \to \mu^+ \Lambda$ and $\bar \nu_\mu p \to \mu^+ \Sigma^0$ reactions. We show the results for $\sigma(E_{\bar \nu_{_\mu}})$ in Fig.~\ref{xsec_enu}, where a
 comparison is made with available experimental results from CERN~\cite{Eichten:1972bb,Erriquez:1977tr,Erriquez:1978pg}, 
 BNL~\cite{Fanourakis:1980si}, FNAL~\cite{Ammosov:1986jn,Ammosov:1986xv} and 
Serpukhov~\cite{Brunner:1989kw} experiments. A reasonable agreement with the experimental results can 
 be seen. We also show in Fig.~\ref{pol_enu}, the energy dependence of averaged polarization components $P_L(E_{\bar \nu_{_\mu}})$ and $P_P(E_{\bar \nu_{_\mu}})$
 for completeness which are defined as~\cite{Graczyk:2004uy}: 
 \begin{equation}\label{energy_pol}
 \langle P_{L,P}(E_{\bar \nu_{_\mu}}) \rangle = \frac{\int^{Q^2_{max}}_{Q^2_{min}} P_{L,P}(Q^2,E_{\bar \nu_{_\mu}})~\frac{d\sigma}{dQ^2}~dQ^2}{\int^{Q^2_{max}}_{Q^2_{min}} \frac{d\sigma}{dQ^2}~dQ^2},
 \end{equation}
 for the processes  $\bar \nu_\mu p \to \mu^+ \Lambda$ and $\bar \nu_\mu n \to \mu^+ \Sigma^-$. It may be observed from Fig.~\ref{pol_enu} that for the process $\bar \nu_\mu p \to \mu^+ \Lambda$, the 
 polarization components $P_L(E_{\bar \nu_{_\mu}})$ and $P_P(E_{\bar \nu_{_\mu}})$ decrease with the increase in energy while for the process $\bar \nu_\mu n \to \mu^+ \Sigma^-$, 
 these polarization components increase with the energy initially and then become almost constant.
\begin{table}
\begin{tabular}{|c|c|c|c|}\hline\hline
& $\langle P_L \rangle$ & $\langle P_P \rangle $\footnote{One may note that, for present work we have considered the sign convention for 
  perpendicular polarization which is opposite to that of used by Erriquez et al.~\cite{Erriquez:1978pg}.} &  $\langle \sigma \rangle$ $\times \;(10^{-40}$ cm$^2)$\\\hline
{\bf Experiments}&&&\\
Erriquez et al.~\cite{Erriquez:1978pg}   & -0.06$\pm$ 0.44 & 1.05 $\pm$ 0.30   & ~~~ 2.07 $\pm$ 0.75~~~~\\
Erriquez et al.~\cite{Erriquez:1977tr}   & -- &  -- &   1.40 $\pm$ 0.41(Propane)\\ 
Eichten et al.~\cite{Eichten:1972bb} &  --    &   --   & 1.3 $\pm ^{0.9}_{0.7}$(Freon) \\ \hline
{\bf Theory}&&&\\
Present work($M_A = 0.84$ GeV)& 0.10&--0.75&2.00\\
($M_A = 1.026$ GeV)&0.05 &  --0.85  & 2.15 \\
($M_A = 1.2$ GeV)&0.03&--0.89&2.31\\
Erriquez et al.~\cite{Erriquez:1978pg}($M_A = 0.84$ GeV)& 0.14 &0.73 &  2.07\\\hline\hline
\end{tabular}
\caption{Flux averaged cross section $\langle \sigma \rangle$(using Eq.~\ref{flux_sigma}), longitudinal $\langle P_L \rangle$ and perpendicular $\langle P_P \rangle$ components of
  polarization(using Eq.~\ref{flux_pol}) are given for the process $\bar \nu_\mu p \to \mu^+ \Lambda$.}
  \label{mema_ff}
\end{table}
\begin{table} 
 \begin{center}
\begin{tabular}{|c|c|c|c|c|c|c|c|}  \hline 
 Spectrum &\multicolumn{2}{|c|}{$<\sigma> \times 10^{-40}~\rm{cm^2}$} & \multicolumn{2}{|c|}{\quad$<P_L>$~~~} & \multicolumn{2}{|c|}{\quad$<P_P>$~~~ }\\ \hline \hline
  &\qquad$\Sigma^-$~~~&$\Lambda$&$~~\Sigma^-~~$&$\Lambda$&$~~\Sigma^-~~$&$\Lambda$\\ \hline

       MicroBooNE~\cite{Chen:2007ae}                  &      0.31&  0.76 &--0.43 & 0.39  &0.37  &--0.78\\   \hline 

       MINER$\nu$A~\cite{Fields:2013zhk}                  &      1.17&  2.5 &--0.42 & --0.03  &0.43  &--0.85\\  \hline 

              T2K~\cite{Abe:2015ibe}          &      0.27&  0.74 &--0.44 & 0.43  &0.37  &--0.75\\  \hline \hline
  \end{tabular}
\end{center}
\caption{Total cross section using Eq.~\ref{flux_sigma}, longitudinal and perpendicular components of polarization using Eq.~\ref{flux_pol} are integrated over various fluxes 
for $\bar{\nu}_\mu(k) + N(p)\rightarrow \mu^+(k^\prime) + Y(p^\prime)$ process using $f_1^{NY}(Q^2),~f_2^{NY}(Q^2),~g_1^{NY}(Q^2)$ 
   from Table-\ref{tab:formfac} and $g_3^{NY}(Q^2)$ from Eq.~\ref{g3_marshak} with $m_\mu \ne 0$ and $M_A$ = 1.026 GeV. }
\label{tab:lambda_tab}
\end{table}
 \subsection{Total cross section and polarizations}
 We have integrated the differential cross section $\frac{d\sigma}{dQ^2}$ and polarization observables $P_L(Q^2)$ and $P_P(Q^2)$ over $E_{\bar \nu_{_\mu}}$ and $Q^2$
 distributions to obtain the total cross section $\langle \sigma \rangle $ defined as: 
 \begin{equation}\label{flux_sigma}
  \langle \sigma \rangle = \frac{\int^{E_{max}}_{E_{th}}\int^{Q^2_{max}}_{Q^2_{min}} \frac{d\sigma}{dQ^2}~dQ^2 \Phi(E_{\bar \nu_{_\mu}}) dE_{\bar \nu_{_\mu}} }
  {\int^{E_{max}}_{E_{min}}\Phi(E_{\bar \nu_{_\mu}}) dE_{\bar \nu_{_\mu}} }
 \end{equation}
 and components of hyperon polarization  $\langle P_{L,P} \rangle$ defined as:
\begin{equation}\label{flux_pol}
\langle P_{L,P} \rangle = \frac{1}{\langle \sigma \rangle} \int_{E_{th}}^{E_{max}} \int^{Q^2_{max}}_{Q^2_{min}} P_{L,P}(Q^2,E_{\bar \nu_{_\mu}})
~ \frac{d\sigma}{dQ^2}~dQ^2 \Phi(E_{\bar \nu_{_\mu}}) d E_{\bar \nu_{_\mu}}.  
\end{equation} 
In order to compare with the experimental results of CERN experiment~\cite{Erriquez:1978pg}, we have performed 
  the numerical calculations for the flux averaged cross section $\langle \sigma \rangle$, longitudinal $\langle P_L 
  \rangle$ and perpendicular $\langle P_P \rangle$ 
  polarization components relevant for the 
  antineutrino flux of SPS antineutrino 
  beam of Gargamelle experiment at CERN~\cite{Armenise:1979zg} and present our results in Table-\ref{mema_ff}. The results are compared with the
  available experimental results from CERN~\cite{Erriquez:1977tr,Eichten:1972bb,Erriquez:1978pg} experiment and the theoretical results quoted 
  by Erriquez et al.~\cite{Erriquez:1978pg}. For reference we also show in Table-\ref{tab:lambda_tab}, our results for $\langle \sigma \rangle$, $\langle P_L \rangle$ and $\langle P_P \rangle$
  relevant for MicroBooNE~\cite{Chen:2007ae}, MINER$\nu$A~\cite{Fields:2013zhk} and T2K~\cite{Abe:2015ibe} experiments, which may be useful in the interpretation
  of the results from these experiments, whenever they become available.

\section{Summary and Conclusions}\label{sec:summary}
We have in this work studied the differential cross section $\frac{d\sigma}{dQ^2}$ as well as  longitudinal($P_L(Q^2)$) and perpendicular ($P_P(Q^2)$) components of polarization 
of $\Lambda$ and $\Sigma$ hyperons produced in the quasielastic reactions
of antineutrinos on free and bound nucleons. The effect of nuclear medium arising due to Fermi motion and Pauli blocking for initial nucleon have been included. The transition form factors for the nucleon-hyperon transition have been obtained using Cabibbo theory
assuming SU(3) invariance and the absence of second class currents. 
 The sensitivity of $Q^2$ dependence on $\frac{d\sigma}{dQ^2}$, $P_L(Q^2)$ and $P_P(Q^2)$ due to the variation in $M_A$ has been studied. The possibility of determining the 
pseudoscalar form factor in $|\Delta S| = 1$ sector has also been explored. The theoretical
results  have been compared with the available experimental results on the energy dependence of the total cross sections from CERN~\cite{Erriquez:1977tr,Eichten:1972bb,Erriquez:1978pg} and other
experiments performed at BNL~\cite{Fanourakis:1980si}, FNAL~\cite{Ammosov:1986jn,Ammosov:1986xv} and Serpukhov~\cite{Brunner:1989kw}. A comparison of our theoretical results with the experimental results
on the flux averaged total cross section and averaged polarization 
components for CERN~\cite{Erriquez:1978pg} experiment has also been made. Predictions for the flux averaged cross section and polarization components  
have been made for the future experiments being done on nuclear targets with antineutrino beams at
 MicroBooNE~\cite{Chen:2007ae}, MINER$\nu$A~\cite{Fields:2013zhk} and T2K~\cite{Abe:2015ibe}.

To summarize our results we find that:
  \begin{enumerate}
  \item The theoretical results for the total cross section as a function of energy i.e. $\sigma(E_{\bar \nu_\mu})$ is found to be in satisfactory agreement with the earlier experimental results 
  available from CERN, BNL and Serpukhov laboratories
  with an axial mass of $M_A$ = 1.026 GeV, the world average value obtained from $\Delta S=0$ experiments.
  \item The longitudinal and perpendicular components of polarization $P_L(Q^2)$ and $P_P(Q^2)$ are sensitive to the value of axial dipole mass $M_A$.
 Therefore, it is possible to determine the value of $M_A$ independent of the cross section measurements for the single hyperon production. 
  \item The $Q^2$ dependence of the cross section $\frac{d\sigma}{dQ^2}$ and  polarization components $P_{L,P}(Q^2)$ are found to be sensitive to the neutron charge form factor in the case of  
  $\bar\nu_\mu n \to \mu^+ \Sigma^-$ process, specially for $Q^2 >$ 0.2 GeV$^2$.
  \item   At lower antineutrino energies $E_{\bar \nu_\mu} \sim 0.5$ GeV, the differential cross section $\frac{d\sigma}{dQ^2}$ and the polarization components $P_{L,P}(Q^2)$ 
  are sensitive to the value of pseudoscalar form factor. It should be possible to test PCAC and GT relation in the strangeness sector, from the quasielastic production of hyperons at lower energies
   relevant to MicroBooNE and T2K experiments. At antineutrino energies $E_{\bar \nu_\mu} \ge 1$ GeV, the differential cross section $\frac{d\sigma}{dQ^2}$ and the polarization components are not found to be sensitive to
  the pseudoscalar form factor.
  \item The effect of nuclear medium on $\frac{d\sigma}{dQ^2}$, $P_L(Q^2)$ and $P_P(Q^2)$ arising due to Fermi motion and Pauli blocking of initial
  nucleon are studied quantitatively. They 
  are found to be quite small and negligible for $\frac{d\sigma}{dQ^2}$. However, these effects are found 
  to be non-negligible but small for $P_L(Q^2)$ and $P_P(Q^2)$ and show no appreciable 
  dependence on the nucleon number A. 
  
  It should be emphasized that we have assumed in our present work the absence  of second class currents.
  If such currents are present, the results are expected to get modified. Moreover, the presence of second class currents will also give rise to T--violating
  effects in quasielastic hyperon production induced by antineutrinos. This work
  is in progress and will be reported in future.

   \end{enumerate}

\section{Appendix}\label{appendix}
 The expressions for ${\cal N}(Q^2,E_{\bar \nu_{_\mu}})$, ${\cal A}(Q^2,E_{\bar \nu_{_\mu}})$ and ${\cal B}(Q^2,E_{\bar \nu_{_\mu}})$  are given as:
 \begin{eqnarray}
{\cal N}(Q^2,E_{\bar \nu_{_\mu}}) &=& f_1^2(2 E_{\bar \nu_{_\mu}} (\vec k \cdot \vec k^{\prime}+2 m_N E_\mu-m_\mu^2)-2 \vec k \cdot \vec k^{\prime} (m_Y+E_\mu))+ \nonumber\\
 && \left. \frac{f_2^2}{(m_N+m_Y)^2}(4 (\vec k \cdot \vec k^{\prime})^2 (m_Y+E_\mu-E_{\bar \nu_{_\mu}})+\vec k \cdot \vec k^{\prime} (m_N (4 (E_\mu^2+E_{\bar \nu_{_\mu}}^2)-m_\mu^2)-\right. \nonumber\\
 &&\qquad \left.3 m_\mu^2 (m_Y+E_\mu-E_{\bar \nu_{_\mu}}))-4 m_N m_\mu^2 E_{\bar \nu_{_\mu}}^2)+\right. \nonumber\\
 && \left. g_1^2(2 (\vec k \cdot \vec k^{\prime} (m_Y-E_\mu+E_{\bar \nu_{_\mu}})-E_{\bar \nu_{_\mu}} (m_\mu^2-2 m_N E_\mu)))+\right. \nonumber\\
 && \left. g_3^2((\vec k \cdot \vec k^{\prime})^2 m_\mu^2 (m_N-m_Y-E_\mu+E_{\bar \nu_{_\mu}})) +\right. \nonumber\\
 && \left. \frac{f_1 f_2}{m_N+m_Y}(8 (\vec k \cdot \vec k^{\prime})^2+\vec k \cdot \vec k^{\prime} (4 (m_N-m_Y) (E_\mu-E_{\bar \nu_{_\mu}})-6 m_\mu^2)+ \right. \nonumber\\
 && \qquad \left. 2 m_\mu^2 E_{\bar \nu_{_\mu}} (m_N-m_Y)) +\right. \nonumber \\
 && \left. f_1 g_1(-4 (\vec k \cdot \vec k^{\prime} (E_\mu+E_{\bar \nu_{_\mu}})-m_\mu^2 E_{\bar \nu_{_\mu}}))+\right. \nonumber\\
 && \left. \frac{f_2 g_1}{m_N+m_Y}(-4 (m_N+m_Y) (\vec k \cdot \vec k^{\prime} (E_\mu+E_{\bar \nu_{_\mu}})-m_\mu^2 E_{\bar \nu_{_\mu}}))+\right. \nonumber\\
 &&  g_1 g_3(-2 m_\mu^2 (\vec k \cdot \vec k^{\prime}+E_{\bar \nu_{_\mu}} (m_Y-m_N)))  \label{j0_factor}
\end{eqnarray}

\begin{eqnarray}
{\cal A}(Q^2,E_{\bar \nu_{_\mu}}) &=& f_1^2 (-2 \vec k \cdot \vec k^{\prime}-(m_N-m_Y) (E_\mu-E_{\bar \nu_{_\mu}})+m_\mu^2) + \nonumber\\
 && \left.\frac{f_2^2}{(m_N+m_Y)^2}((2 \vec k \cdot \vec k^{\prime}-m_\mu^2) (2 \vec k \cdot \vec k^{\prime}+(m_N-m_Y) (E_\mu-E_{\bar \nu_{_\mu}})-m_\mu^2)) + \right. \nonumber\\
 && \left.g_1^2(2 \vec k \cdot \vec k^{\prime}+(m_N+m_Y) (E_\mu-E_{\bar \nu_{_\mu}})-m_\mu^2) + \right. \nonumber\\
 && \left.\frac{f_1 f_2}{m_N+m_Y}(-2 (2 \vec k \cdot \vec k^{\prime} (m_Y+E_\mu-E_{\bar \nu_{_\mu}})+m_N (E_\mu-E_{\bar \nu_{_\mu}})^2 +\right. \nonumber \\
 && \qquad \left. m_\mu^2 (-(m_Y+E_\mu-E_{\bar \nu_{_\mu}})))) +\right. \nonumber\\
 && \left.f_1 g_1(2 m_Y (E_\mu+E_{\bar \nu_{_\mu}})) +
       f_1 g_3(m_\mu^2 (-m_N+m_Y+E_\mu-E_{\bar \nu_{_\mu}})) +\right. \nonumber\\
 && \left.\frac{f_2 g_1}{m_N+m_Y}(-4 \vec k \cdot \vec k^{\prime} (E_\mu+E_{\bar \nu_{_\mu}})+m_N (m_\mu^2-2 E_\mu^2+2 E_{\bar \nu_{_\mu}}^2)+m_\mu^2 (m_Y+E_\mu+3 E_{\bar \nu_{_\mu}})) +\right. \nonumber\\
 && \frac{f_2 g_3}{m_N+m_Y}(m_\mu^2 (-2 \vec k \cdot \vec k^{\prime}-(m_N-m_Y) (E_\mu-E_{\bar \nu_{_\mu}})+m_\mu^2)),\label{a_factor}
\end{eqnarray}

\begin{eqnarray}
{\cal B}(Q^2,E_{\bar \nu_{_\mu}}) &=& f_1^2((E_\mu+E_{\bar \nu_{_\mu}}) (2 \vec k \cdot \vec k^{\prime}+m_Y (m_Y-m_N))+m_\mu^2 (m_Y-2 E_{\bar \nu_{_\mu}})) + \nonumber\\
 && \left.\frac{f_2^2}{(m_N+m_Y)^2}(4 (\vec k \cdot \vec k^{\prime})^2 (E_\mu+E_{\bar \nu_{_\mu}})+2 \vec k \cdot \vec k^{\prime} ((E_\mu+E_{\bar \nu_{_\mu}})(m_N (m_Y+2 E_\mu-2 E_{\bar \nu_{_\mu}})+\right. \nonumber\\
 && \qquad\left.m_Y^2)-m_\mu^2 (m_Y+E_\mu+3 E_{\bar \nu_{_\mu}}))+m_\mu^2 (-m_N (m_Y(E_\mu+E_{\bar \nu_{_\mu}})+4 E_{\bar \nu_{_\mu}} (E_\mu-E_{\bar \nu_{_\mu}}))+\right. \nonumber\\
 && \qquad\left.m_\mu^2 (m_Y+2 E_{\bar \nu_{_\mu}})+m_Y^2 (E_\mu-3 E_{\bar \nu_{_\mu}}))) + \right. \nonumber\\
 && \left.g_1^2((E_\mu+E_{\bar \nu_{_\mu}}) (2 \vec k \cdot \vec k^{\prime}+m_Y (m_N+m_Y))-m_\mu^2 (m_Y+2 E_{\bar \nu_{_\mu}})) +\right. \nonumber\\
 && \left.\frac{f_1 f_2}{m_N+m_Y}(2 (m_N (E_\mu+E_{\bar \nu_{_\mu}}) (2 \vec k \cdot \vec k^{\prime}+m_Y (E_{\bar \nu_{_\mu}}-E_\mu))+m_\mu^2 (m_Y (m_Y+E_\mu)-\right. \nonumber\\
 && \qquad \left. E_{\bar \nu_{_\mu}} (2 m_N+m_Y)))) + \right. \nonumber\\
 && \left.f_1 g_1(2 E_\mu (2 \vec k \cdot \vec k^{\prime}+m_Y^2)-2 E_{\bar \nu_{_\mu}} (2 \vec k \cdot \vec k^{\prime}+4 m_N E_\mu-2 m_\mu^2+m_Y^2)) +\right. \nonumber\\
 && \left.f_1 g_3(m_\mu^2 (2 \vec k \cdot \vec k^{\prime}-m_N (m_Y+2 E_{\bar \nu_{_\mu}})+m_Y (m_Y+E_\mu-E_{\bar \nu_{_\mu}}))) +\right. \nonumber\\
 && \left.\frac{f_2 g_1}{m_N+m_Y}(-8 (\vec k \cdot \vec k^{\prime})^2+\vec k \cdot \vec k^{\prime} (6 m_\mu^2-4 (m_N E_\mu-m_N E_{\bar \nu_{_\mu}}+m_Y^2))\right. \nonumber\\
 && \qquad \left. +m_N (m_\mu^2 (m_Y-2 E_{\bar \nu_{_\mu}})-
     2 m_Y (E_\mu+E_{\bar \nu_{_\mu}})^2)+m_\mu^2 m_Y (m_Y+E_\mu+3 E_{\bar \nu_{_\mu}})) +\right. \nonumber\\
 && \frac{f_2 g_3}{m_N+m_Y}(m_\mu^2 ((E_\mu+E_{\bar \nu_{_\mu}}) (2 \vec k \cdot \vec k^{\prime}+m_Y (m_Y-m_N))+m_\mu^2 (m_Y-2 E_{\bar \nu_{_\mu}}))).\label{b_factor}
\end{eqnarray}

\begin{eqnarray}
 \alpha(Q^2,\vec{p})&=& \frac{64}{m_Y}\left[f_1^2\left({k \cdot k^\prime} {k \cdot p}-m_N m_Y \left({k \cdot k^\prime}+{k^{\prime} \cdot p}
 -m_\mu^2\right)+{k \cdot k^\prime} {k^{\prime} \cdot p}-{k \cdot p} m_\mu^2+{k^{\prime} \cdot p} m_Y^2 \right)\right.\nonumber\\ 
 &&\qquad +\left. \frac{f_2^2}{(m_N+m_Y)^2} \left(2 {k \cdot k^\prime}^2 ({k \cdot p}+{k^{\prime} \cdot p}+m_N m_Y)-{k \cdot k^\prime} \left(2 {k \cdot p}^2+3 {k \cdot p} m_\mu^2-2 {k \cdot p}
 m_Y^2-2 {k^{\prime} \cdot p}^2 \right. \right. \right. \nonumber\\
 && \qquad\left. \left. \left.-2 {k^{\prime} \cdot p} m_N m_Y+{k^{\prime} \cdot p} m_\mu^2+3 m_N m_\mu^2 m_Y\right)+m_\mu^2 \left(2 {k \cdot p}^2+{k \cdot p}
 \left(-2 {k^{\prime} \cdot p}+m_\mu^2-2 m_Y^2\right)\right. \right. \right. \nonumber\\
 && \qquad\left. \left.\left.+m_Y \left(-{k^{\prime} \cdot p} m_N+{k^{\prime} \cdot p} m_Y+m_N m_\mu^2\right)\right) \right)\right. \nonumber \\
 && \left.g_1^2 \left({k \cdot k^\prime} ({k \cdot p}+{k^{\prime} \cdot p}+m_N m_Y)-{k \cdot p} m_\mu^2+m_Y \left({k^{\prime} \cdot p} (m_N+m_Y)-m_N m_\mu^2\right) \right)\right. 
 \nonumber \\ 
 && \left.\frac{f_1 f_2}{m_N+m_Y} \left(2 \left({k \cdot k^\prime} m_N ({k \cdot p}+{k^{\prime} \cdot p})+m_Y ({k \cdot p}-{k^{\prime} \cdot p}) \left({k \cdot k^\prime}+{k^{\prime} \cdot p}-m_\mu^2\right)
 \right.\right.\right.\nonumber \\
 &&\qquad \left.\left.\left.+m_N m_Y^2 
 \left(m_\mu^2-{k \cdot k^\prime}\right)-{k \cdot p} m_N m_\mu^2\right) \right)\right.\nonumber \\ 
 && +\left.f_1 g_1\left(2 \left({k \cdot k^\prime} ({k^{\prime} \cdot p}-{k \cdot p})+{k \cdot p} \left(m_\mu^2-2 {k^{\prime} \cdot p}\right)+{k^{\prime} \cdot p} m_Y^2\right) \right)\right. \nonumber\\
 && +\left.f_1 g_3 \left(m_\mu^2 ({k \cdot k^\prime} m_N-{k \cdot p} (m_N+m_Y)+m_Y ({k^{\prime} \cdot p}+m_N m_Y-m_N)) \right) \right. \nonumber\\ 
 && \left.\frac{f_2 g_1}{m_N+m_Y}\left(-4 {k \cdot k^\prime}^2 m_N+{k \cdot k^\prime} \left(m_N \left(2 {k \cdot p}-2 {k^{\prime} \cdot p}+3 m_\mu^2\right)-2 m_Y ({k \cdot p}
 +{k^{\prime} \cdot p})-2 m_N m_Y^2\right) \right.\right.\nonumber \\ 
 && \qquad\left.-m_\mu^2 ({k \cdot p} (m_N-3 m_Y)-m_Y ({k^{\prime} \cdot p}+m_N m_Y+m_N))-2 {k^{\prime} \cdot p} m_Y
 ({k \cdot p}+{k^{\prime} \cdot p}) \right) \nonumber\\
 && \left.+ \frac{f_2 g_3}{m_N+m_Y}\left(m_\mu^2 \left({k \cdot k^\prime} {k \cdot p}-m_N m_Y \left({k \cdot k^\prime}+{k^{\prime} \cdot p}-m_\mu^2\right)+{k \cdot k^\prime} 
 {k^{\prime} \cdot p}-{k \cdot p} m_\mu^2+{k^{\prime} \cdot p} m_Y^2\right) \right)\right]\nonumber\\
\end{eqnarray}

\begin{eqnarray}
 \beta(Q^2,\vec{p}) &=&\frac{64}{m_Y}\left[f_1^2 \left({k \cdot p} \left(m_Y (m_N-m_Y)+m_\mu^2\right)-{k \cdot k^\prime} ({k \cdot p}+{k^{\prime} \cdot p}+
 m_N m_Y) \right)\right.\nonumber\\
 && \left. \frac{f_2^2}{(m_N+m_Y)^2} \left( -2 {k \cdot k^\prime}^2 ({k \cdot p}+{k^{\prime} \cdot p}-m_N m_Y)+{k \cdot k^\prime} \left(2 {k \cdot p}^2-m_N m_Y \left(2 {k \cdot p}+
 m_\mu^2\right)\right.\right.\right.\nonumber\\
 && \qquad\left. \left. +3 {k \cdot p} m_\mu^2-2 {k^{\prime} \cdot p}^2+{k^{\prime} \cdot p} m_\mu^2-2 {k^{\prime} \cdot p} m_Y^2\right)+{k \cdot p} m_\mu^2 \left(-2 {k \cdot p}+2 {k^{\prime} \cdot p}+
 m_Y (m_N+m_Y)-m_\mu^2\right)\right)\nonumber\\
 &&\left.g_1^2\left( {k \cdot p} \left(m_\mu^2-m_Y (m_N+m_Y)\right)-{k \cdot k^\prime} ({k \cdot p}+{k^{\prime} \cdot p}-m_N m_Y)\right)\right.\nonumber\\
 && \left.\frac{f_1 f_2}{m_N+m_Y}\left(-2 \left({k \cdot k^\prime} m_N ({k \cdot p}+{k^{\prime} \cdot p})-m_Y ({k \cdot k^\prime}-{k \cdot p}) ({k \cdot p}-{k^{\prime} \cdot p})+{k \cdot k^\prime} m_N
 m_Y^2-{k \cdot p} m_N m_\mu^2\right) \right)\right.\nonumber\\
 && \left.f_1 g_1 \left(2 \left({k \cdot k^\prime} ({k \cdot p}-{k^{\prime} \cdot p})+{k \cdot p} \left(2 {k^{\prime} \cdot p}-m_\mu^2+m_Y^2\right)\right)\right)\right.\nonumber\\
 &&\left. f_1 g_3\left(m_N m_\mu^2 ({k \cdot p}-{k \cdot k^\prime}) \right)\right.\nonumber\\
 && \left.\frac{f_2 g_1}{m_N+m_Y}\left(m_N \left(4 {k \cdot k^\prime}^2+{k \cdot k^\prime} \left(-2 {k \cdot p}+2 {k^{\prime} \cdot p}-3 m_\mu^2\right)+{k \cdot p} m_\mu^2\right)\right.\right.\nonumber \\
 &&\qquad-2 \left.m_Y({k \cdot k^\prime}-{k \cdot p}) ({k \cdot p}+{k^{\prime} \cdot p})+2 {k \cdot k^\prime} m_N m_Y^2 \right)\nonumber\\
 &&\left.\frac{f_2 g_3}{m_N+m_Y} \left( m_\mu^2 \left({k \cdot p} \left(m_Y (m_N-m_Y)+m_\mu^2\right)-{k \cdot k^\prime} ({k \cdot p}+{k^{\prime} \cdot p}+m_N
 m_Y)\right)\right)\right]
\end{eqnarray}

\begin{eqnarray}
 \eta(Q^2,\vec{p})&=& \frac{64}{m_Y}\left[ f_1^2\left({k \cdot k^\prime} ({k \cdot p}+{k^{\prime} \cdot p})-{k \cdot p} m_\mu^2 \right)\right.\nonumber\\
 &&\left.\frac{f_2^2}{(m_N+m_Y)^2} \left(\left(2 ({k \cdot k^\prime}-{k \cdot p}+{k^{\prime} \cdot p})-m_\mu^2\right) \left({k \cdot k^\prime}
 ({k \cdot p}+{k^{\prime} \cdot p})-{k \cdot p} m_\mu^2\right)\right)\right.\nonumber\\
 &&\left.g_1^2\left({k \cdot k^\prime} ({k \cdot p}+{k^{\prime} \cdot p})-{k \cdot p} m_\mu^2 \right)\right.\nonumber\\
 &&\left.\frac{f_1 f_2}{m_N+m_Y}\left(2 m_N \left({k \cdot k^\prime} ({k \cdot p}+{k^{\prime} \cdot p})-{k \cdot p} m_\mu^2\right)\right)\right.\nonumber\\
 &&\left.f_1 g_1\left(2 {k \cdot k^\prime} ({k^{\prime} \cdot p}-{k \cdot p})+2 {k \cdot p} \left(m_\mu^2-2 {k^{\prime} \cdot p}\right)\right)\right.\nonumber\\
 &&\left.f_1 g_3\left(m_\mu^2 ({k \cdot k^\prime}-{k \cdot p}) (m_N-m_Y) \right)\right.\nonumber\\
 &&\left.\frac{f_2 g_1}{m_N+m_Y}\left(-4 {k \cdot k^\prime}^2 m_N+{k \cdot k^\prime} \left(2 {k \cdot p} m_N-2 {k^{\prime} \cdot p} m_N\right.\right.\right.\nonumber\\
 &&\qquad+\left.\left.\left.m_\mu^2 (3 m_N+m_Y)\right)-{k \cdot p} m_\mu^2 (m_N+m_Y)\right)\right.\nonumber\\
 &&\left.\frac{f_2 g_3}{m_N+m_Y}\left({k \cdot k^\prime} m_\mu^2 ({k \cdot p}+{k^{\prime} \cdot p})-{k \cdot p} m_\mu^4\right)\right]
\end{eqnarray}

\section{Acknowledgment}   
M. S. A. is thankful to Department of Science and Technology(DST), Government of India for providing
financial assistance under Grant No. SR/S2/HEP-18/2012.

 \end{document}